# Optimal Auctions with Correlated Bidders are Easy

Shahar Dobzinski
Department of Computer Science
Cornell Unversity
shahar@cs.cornell.edu

Hu Fu
Department of Computer Science
Cornell Unversity
hufu@cs.cornell.edu

Robert Kleinberg
Department of Computer Science
Cornell Unversity
rdk@cs.cornell.edu

October 8, 2018

**Abstract**

We consider the problem of designing a revenue-maximizing auction for a single item, when the values of the bidders are drawn from a correlated distribution. We observe that there exists an algorithm that finds the optimal randomized mechanism that runs in time polynomial in the size of the support. We leverage this result to show that in the oracle model introduced by Ronen and Saberi [FOCS'02], there exists a polynomial time truthful in expectation mechanism that provides a $(\frac 3 2 + \epsilon)$-approximation to the revenue achievable by an optimal truthful-in-expectation mechanism, and a polynomial time deterministic truthful mechanism that guarantees $\frac 5 3$ approximation to the revenue achievable by an optimal deterministic truthful mechanism.

We show that the $\frac 5 3$-approximation mechanism provides the same approximation ratio also with respect to the optimal truthful-in-expectation mechanism. This shows that the performance gap between truthful-in-expectation and deterministic mechanisms is relatively small. En route, we solve an open question of Mehta and Vazirani [EC'04].

Finally, we extend some of our results to the multi-item case, and show how to compute the optimal truthful-in-expectation mechanisms for bidders with more complex valuations.

# 1  Introduction

Myerson, in his seminal paper [9], studies the following problem: $n$ bidders are competing on a single item. Each bidder's value for the item is drawn independently from a distribution. What is the optimal revenue-maximizing auction? Myerson gives a complete and simple characterization of the optimal auctions. A natural open question raised by this work is to analyze the case when the bidders' values are drawn from a general (correlated) distribution. That question is the topic of this paper.

Naturally, the problem has been heavily studied in economics. Usually, the economics approach involved attempts to characterize restricted special cases. See [5] for a survey. One notable exception was given by Cremer and McLean [4], who show that sometimes an auction that extracts the *full social welfare* exists. However, the proposed solution works only for a restricted class of distributions. More severely, although each bidder's expected gain from participating in the mechanism is zero, it is common for a bidder to be charged an amount greatly exceeding her value. Hence the bidders have little incentive to participate in the auction in the first place.

Ronen [12] offers a thought-provoking alternative to the economics approach: instead of struggling to find an exact characterization of optimal auctions – even though a simple characterization probably does not exist in the case of correlated distributions – we should design auctions that are *approximately* optimal. Ronen presents the following elegant truthful mechanism, the *lookahead auction*, that provides in expectation at least half of the revenue of the optimal auction. First, find the $n-1$ bidders with the lowest values. Given their values, consider the conditional distribution of the value $v_i$ of bidder $i$ with the highest valuation, and calculate a price $p$ that maximizes the expected revenue. If $v_i \geq p$, bidder $i$ is assigned the item and is charged $p$, otherwise no one is assigned the item and no one pays anything.

We continue the line of research of studying approximately optimal auctions. In particular, we investigate the following three research directions.

**Research Direction I: Obtaining Better Approximation Ratios**

Perhaps the first question every theoretical computer scientist would ask following Ronen's paper is: "Are there mechanisms with better approximation ratios?" To answer this question we have to be more explicit about the computational model in hand. Ronen and Saberi [13] introduce the *oracle model*: the distribution is given to us as a black box and we are allowed to ask *conditional-distribution* queries. That is, given the values of $n-k$ players, what is the conditional distribution of the values of the remaining $k$ players? Ronen and Saberi prove several hardness results in this model, in particular that no *ascending* auction can guarantee a $\frac{4}{3}$-approximation. The other model is the *explicit model*, where the running time has to be polynomial in the support size of the distribution. Papadimitriou and Pierrakos [11] prove that it is possible to exactly compute the optimal deterministic auction for 2 bidders, but NP-hard to do so for more bidders.

**Research Direction II: Relaxing the Solution Concept**

This paper considers three extensively studied notions of truthfulness. The simplest and strongest one is *dominant strategy* truthfulness (or deterministic truthfulness): a profit-maximizing strategy of each bidder is to reveal his true value. The second notion is *randomized universal truthfulness*: here a mechanism is a probability distribution over deterministic truthful mechanisms. The third notion is that of *truthfulness in expectation*: revealing the true value maximizes the *expected* profit of each bidder, where the expectation is taken over the internal random coins of the mechanism.

Myerson [9] shows that the optimal auction is always deterministic when bids are independent, even if truthfulness in expectation is allowed[1]. We pose the following question: can truthful-in-expectation mechanisms achieve more revenue than deterministic mechanisms when the distribution is correlated?

---

[1]In fact, he proves that this is true even under the weaker notion of Bayesian truthfulness (see [10] for a definition).

If so, how much more? Notice that this question is of interest as a pure existence question, though it is also interesting to consider the refinement in which the mechanisms are required to be computationally efficient.

**Research Direction III: Beyond Single-Item Auctions**

Naturally, we would like to design revenue-maximizing auctions also for more complicated settings. The direct characterization approach of economists was successful so far only in limited settings — for example, Armstrong [1] characterizes optimal auctions for two items and two bidders with additive valuations, where each bidder has only two possible values for each item and the valuations are independently distributed — while the approximation approach of computer scientists was mostly successful when considering the weak solution concept of Bayesian truthfulness [3, 2][2].

**Our Results**

Our paper contributes to all three research directions sketched above while drawing connections between them. The starting point of our investigations is one, quite simple, observation: in the explicit model the optimal truthful-in-expectation auction can be computed in polynomial time. As we explain in Section 3, this auction can be found by solving a natural linear program[3] that encodes allocation probabilities, expected payments, and the incentive constraints that link them. Unlike the mechanism of Cremer and McLean [4], ours never charges agents an amount greater than their bid value, a property shared by all of the mechanisms we construct in this paper.

Unfortunately, while the explicit model might be useful for settings with a small number of players, for large numbers of players the distribution usually has exponentially large support, e.g. when each player has two possible values and valuations are independent. We overcome this obstacle in Section 5 by reducing the optimal auction design problem for any number of players to the problem of designing optimal auctions for a constant number of players, for which the LP-based approach is feasible. Consider the following extension of the lookahead auction, termed the *$k$-lookahead auction.* Find the $n-k$ bidders with the lowest values. Given their values, consider the conditional distribution of the values of the $k$ bidders with the highest valuations, and run the optimal auction for these $k$ bidders[4]. We show that this auction is a $\frac{3k-1}{2k-1}$-approximation to the optimal revenue[5]. In particular, this proves that *there exists a polynomial time $(\frac{3}{2} + \epsilon)$-approximation truthful-in-expectation mechanism in the oracle model.*

In Section 4 we consider deterministic truthful mechanisms. We show the following general "de-randomization" result: *for every truthful-expectation mechanism for 2 bidders and a single item, there exists a universally truthful mechanism with the same allocation function and the same payments.* In other words, relaxing the solution concept to truthfulness in expectation is useless, as every mechanism (in the above setting) can be implemented using the stronger notion of universal truthfulness.

This result has several implications. The first one: *for two players, the optimal deterministic mechanism has the same revenue as the optimal truthful-in-expectation mechanism, and can be found in polynomial time.* We achieve this by computing an optimal truthful-in-expectation mechanism $A$ for two bidders using the LP approach, and then "derandomizing" it into a universally truthful algorithm $A'$ with the same expected revenue. Now, since all (deterministic) mechanisms in the support of $A'$ must have the same expected revenue with respect to the distribution (otherwise $A'$ is not optimal), every one of them is an optimal revenue-maximizing deterministic mechanism.

---

[2]Notice that in our setting we do not even assume the bidders are aware of the existence of an underlying distribution.

[3]Although some caution is needed as we have to define the mechanism for all possible values in the domain, not just for values in the support of the distribution.

[4]This auction may be either deterministic or truthful-in-expectation, depending on the setting.

[5]Ronen [12] claims that the approximation ratio of the $k$-lookahead auction is no better than 2, for every $k$, but his proof is incorrect. He essentially claims (without a proof) that all truthful mechanisms achieve an expected revenue of at most 1 for a certain distribution for 2 bidders. However, the pivot auction we define in Section 5 provides an expected revenue of 1.5 for that distribution.

Another implication is that the 2-lookahead auction is in fact deterministic, even if truthfulness-in-expectation is allowed. We thus obtain the result that *there exists a **deterministic** polynomial time truthful mechanism in the oracle model that guarantees $\frac{3}{5}$-fraction of the revenue of the optimal truthful-in-expectation mechanism*, and, as a corollary, that *for every truthful-in-expectation mechanism with an expected revenue of $r$, there exists a deterministic mechanism with expected revenue of at least $\frac{3}{5}r$.*

We stress that our derandomization result holds regardless of the objective function and even in the prior free setting. We complement this result by showing that if each player has only two possible values for the item ("low" and "high") then the mechanism can be implemented as a universally truthful mechanism, for any number of players. On the other hand, there is a 3-bidder 3-value truthful-in-expectation mechanism that cannot be implemented as a universally truthful algorithm. Together, this answers an open question of Mehta and Vazirani [8].

We show that our linear programming technique extends to yield computationally efficient, optimal truthful-in-expectation mechanisms for a number of settings beyond single-item auctions. First, we generalize to arbitrary single-parameter domains, showing that our linear program constitutes an efficient reduction from revenue maximization to social welfare maximization, in the explicit model when bidders have correlated types. The existence of such an efficient reduction in the case of independent type distributions with polynomially bounded support size was known from the work of Myerson [9].

Next, we show that our technique also extends to certain multi-parameter domains: unit-demand valuations and additive valuations, again yielding computationally efficient optimal truthful-in-expectation mechanisms in the explicit model. The key to these results lies in "decomposing" the fractional solution as a convex combination of integer solutions. For the additive case we present a direct decomposition, while for the-unit demand case we use the classical Birkhoff-von Neumann Theorem. We complement these positive results with a negative one: it is NP-hard to design optimal truthful-in-expectation mechanisms for bidders with OXS valuations, in contrast to the social welfare maximization problem which is computationally easy for this class of valuations [6]. This section is postponed to the appendix (Section A) due to lack of space.

## Open Questions

While this paper studies several old questions, it also raises some new ones. Let us mention a few. In the single-item auction setting, what is the best approximation ratio that can be obtained in polynomial time by deterministic mechanisms? And by truthful-in-expectation mechanisms? We know the answer only for truthful-in-expectation mechanisms in the explicit model, and have gaps in all other cases. The key for a solution might be a better analysis of the $k$-lookahead auction. Our best lower bound on the approximation ratio of the $k$-lookahead auction is $\frac{k+1}{k}$. Does the approximation ratio of the $k$-lookahead auction approach 1 as $k$ grows? We suspect that it does not, but have been unable to prove it.

Another question is to understand how well deterministic mechanisms perform compared to their truthful-in-expectation counterparts. We showed that the ratio is at most $\frac{5}{3}$. In Appendix D we present an example demonstrating that the ratio is at least 1.001. Quantifying the gap essentially boils down to the purely combinatorial problem of analyzing the *integrality gap* of our linear program.

We also provided conditions in which truthful-in-expectation mechanisms can be implemented as universally truthful mechanisms. Can this be extended to other single-parameter settings, such as scheduling on related machines?

The multi-item setting also has plenty of questions to offer. We showed that for additive and unit demand valuations the optimal truthful-in-expectation mechanism can be efficiently computed in the explicit model, but that it is hard to do so for the more general OXS class. Can we design truthful mechanisms with a good approximation ratio for the OXS class, and for the richer gross substitutes class?

The question is of interest also in the oracle model. Also, to what extent do truthful-in-expectation mechanisms outperform deterministic mechanisms in these settings, ignoring computational issues?

Finally, it would be very interesting to study other ways of specifying the type distribution in both the single and multi item settings. Specifically, in what cases can good approximations be obtained in polynomial time, if we are only given black-box access to samples from the distribution?

**Relation to the work of Papadimitriou and Pierrakos**

An unpublished paper of Papadimitriou and Pierrakos [11] is highly related to ours. Their work establishes that optimal deterministic single-item auctions can be computed in polynomial time for two bidders with correlated distributions, but the corresponding problem is NP-hard with three or more bidders. One consequence of our approach of linear programming combined with derandomization is an alternative proof of one of their results, namely the existence of a polynomial-time algorithm to compute the optimal two-bidder deterministic auction. However, they are the first discoverers of this result; we obtained our proof only after learning that they had such an algorithm.

Aside from this one result, the two papers are largely complementary. Their work focuses on deterministic mechanisms, small numbers of players, and computational complexity; ours emphasizes randomized mechanisms, large numbers of players, and approximation.

## 2 Preliminaries

**The Single-Item Setting**

We have one item and $n$ bidders, where bidder $i$ has a privately known value $v_i \geq 0$. We assume some distribution $\mathcal{D}$ on the values of bidders. We allow the distribution of different bidders' values to be correlated. A mechanism $M$ takes a bid vector $v$ and returns an allocation and a price for each bidder. We use $M(v) = (x_1, x_2, \ldots, x_n)$ to denote the allocation vector. In randomized mechanisms, we allow allocations to be fractions. For example, $M(v)_i = x_i$ is the probability with which bidder $i$ gets the item when the bid vector is $v$. Alternatively, we say bidder $i$ gets a $M(v)_i$ fraction of the item. When the mechanism is deterministic, each $x_i$ has to be 0 or 1. We require that $\sum_i x_i \leq 1$ in any allocation vector. This condition is called the *feasibility* of $M$. In *single-parameter domains*, the feasibility condition is generalized by stipulating that $M(v)$ must lie in a specified set of feasible vectors or, in the case of randomized mechanisms, the convex hull of the feasible vectors. When bidder $i$'s bid is $v_i$ and he gets a fraction $x_i$ of the item, the expected payment he makes is required to be at most $v_i x_i$. This is the *individual rationality (IR)* condition. A mechanism is *ex-post IR* if, with any outcome of the random coins, the payment made by the bidder who gets the item is at most his bid. Our goal is to maximize the total expected payment (the revenue) of the bidders, where expectation is taken over $\mathcal{D}$.

The valuations of all bidders but bidder $i$ are denoted by $v_{-i}$. Fix $v_{-i}$. An auction (equivalently, a mechanism) is called *truthful* if for every $v_i, v'_i$, we have $x_i \cdot v_i - p_i \geq x'_i \cdot v_i - p'_i$. Here $x_i, p_i$ (resp. $x'_i, p'_i$) denote the allocation and expected payment when bidder $i$ declares a value of $v_i$ (resp. $v'_i$). Note that when $x_i$ and $x'_i$ can be fractions, the two sides of the inequality are the *expected* profits of bidder $i$ if he bids $v_i$ and $v'_i$, respectively. Therefore the mechanism can be called *truthful in expectation*: truthful for bidders that maximize their expected utility. A mechanism is *universally truthful* if it is a probability distribution over deterministic truthful mechanisms. Notice that the distribution $\mathcal{D}$ is not involved in our definitions of truthfulness, i.e., our mechanisms are truthful for all possible values, not just values in the support of $\mathcal{D}$. In particular, observe that bidders do not need to have any knowledge of $\mathcal{D}$, or any other common prior.

For a distribution $\mathcal{D}$ and a mechanism $M$, let $E_\mathcal{D}[M]$ denote the revenue of $M$. A mechanism $M$ is called an $\alpha$-approximation if $\frac{E_\mathcal{D}[OPT]}{E_\mathcal{D}[M]} \leq \alpha$, where OPT is the revenue-maximizing truthful mechanism given $\mathcal{D}$. (OPT might be truthful or truthful in expectation, depending on the setting.)

To determine the running time of the mechanism, we use two different models. In the *explicit model* we get an explicit list of every type in the support of $\mathcal{D}$ and its probability. The auction should run in time polynomial in $n$ and the description of $\mathcal{D}$. In the *oracle model*, introduced by Ronen and Saberi [13], the distribution is represented by an oracle that can answer the following type of queries[6]: $\Pr[a_1 = v_1, \ldots, a_n = v_n]$, which may include conditional distribution queries. In this model the running time should be polynomial in $n$ and in the combined support size of the players' marginal distributions (i.e., the quantity $\Sigma_i \#\{v_i \mid \exists v_{-i} \text{ such that } (v_i, v_{-i}) \in \text{support}(\mathcal{D})\}$). We assume that the support of each marginal distribution is known to the mechanism in advance.

**Multi-Item Settings**

In the last part of the paper we discuss multi-item issues. Here we have $m$ items and $n$ bidders, where bidder $i$ has a valuation function $v_i$ that gives a non-negative value for each bundle $S$ of items. We will consider several restrictions on the valuation functions, but we define them in the relevant sections. The definitions from the single item case extend naturally: each bidder $i$ is assigned a bundle $S_i$ (or a distribution over bundles in the randomized case) and charged $p_i$, where $p_i = 0$ in case $S_i$ is the empty bundle. We assume a distribution. As before, we want to maximize the expected revenue.

A mechanism is truthful if each bidder maximizes his profit by revealing his true valuation, and is truthful-in-expectation if revealing his true valuation maximizes his expected profit. The definition of approximation ratio extends naturally. In this paper, for multi-item settings we are interested only in the explicit model, where the support of the distribution is explicitly enumerated. The running time of mechanisms should be polynomial in $m$, $n$, and the size of the support.

## 3 An Optimal Truthful in Expectation Mechanism

In this section we show that the optimal truthful in expectation mechanism can be computed in time polynomial in the size of the distribution $\mathcal{D}$ (the explicit model). We leverage this result later: in Section 5 we analyze the approximation ratio of the $k$-lookahead auction. To make the $k$-lookahead auction efficient, the construction of this section is involved as a subroutine. As a consequence of Section 4 we have that the optimal truthful-in-expectation mechanism for two bidders is deterministic. Finally, in Appendix A we extend this construction to several more complicated domains.

The result of this section is based on writing a natural linear program for the problem. We have to be a bit careful when interpreting the linear program as a truthful mechanism, as the mechanism has to be defined over all possible bid vectors, not just those in the support of $\mathcal{D}$. Towards this end, we will need some notation and definitions. Let $\mathcal{D}_i = \{v_i \mid \vec{v} \in \mathcal{D}\}$. We assume that for every $i$, $0 \in \mathcal{D}_i$. This is without loss of generality since we may assume that $\Pr_\mathcal{D}[(0, \vec{v}_{-i})] = 0$. Let $\mathcal{T} = \{(d_i, v_{-i}) \mid d_i \in \mathcal{D}_i \text{ and } v_{-i} \in \mathcal{D}_{-i}\}$. Observe that $|\mathcal{T}| \leq n \cdot |\mathcal{D}|^2$.

- Solve the following linear program:

  *Maximize:* $\sum_{\vec{v} \in \mathcal{D}} \Pr_\mathcal{D}[\vec{v}] \sum_i p_{i,\vec{v}}$

  *Subject to:*

    – For each $\vec{v} \in \mathcal{T}$: $\sum_{i,\vec{v}} x_{i,\vec{v}} \leq 1$.
    – For each bidder $i, \vec{v} \in \mathcal{T}, v'_i \in \mathcal{D}_i$: $x_{i,\vec{v}} \cdot v_i - p_{i,\vec{v}} \geq x_{i,(v'_i, \vec{v}_{-i})} \cdot v_i - p_{i,(v'_i, \vec{v}_{-i})}$.
    – For each $i, \vec{v} \in \mathcal{T}$: $x_{i,\vec{v}} \geq 0, p_{i,\vec{v}} \geq 0$.
    – For each $i, \vec{v}_{-i} \in \mathcal{T}$: $p_{i,(0,\vec{v}_{-i})} = 0$.

---

[6]The oracle used in this paper is strictly weaker than the oracle of [13]. Since our interest in the oracle model lies in designing efficient algorithms, this only strengthen our result.

- Let $\vec{v} = v_1, \ldots, v_n$ be the (realized) valuations of the bidders. The allocation and payments are determined according to the following cases:

  1. If $\vec{v} \in \mathcal{D}$, then allocate the item to exactly one bidder, where each bidder $i$ receives the item with probability $x_{i,\vec{v}}$. Let the payment of the winning bidder $i$ be $\frac{p_{i,\vec{v}}}{x_{i,\vec{v}}}$. The other bidders pay 0.

  2. If $\vec{v} \notin \mathcal{D}$, every bidder $i$ for which $v_{-i} \notin \mathcal{D}_{-i}$ is not allocated the item and does not pay anything. For each bidder $i$ for which $\vec{v}_{-i} \in \mathcal{D}_{-i}$, let $v'_i = \arg\max_{v'_i \in \mathcal{D}_i} x_{i,(v'_i, v_{-i})} \cdot v_i - p_{i,(v'_i, v_{-i})}$. Each such bidder $i$ receives the item with probability $x_{i,(v'_i, v_{-i})}$ and pays $\frac{p_{i,(v'_i, v_{-i})}}{x_{i,(v'_i, v_{-i})}}$ contingent upon receiving the item.

**Theorem 3.1** *The mechanism is feasible, individually rational, truthful in expectation, and runs in time polynomial in the size of the distribution $\mathcal{D}$. Its expected revenue is equal to the expected revenue of the optimal (individually rational) truthful in expectation mechanism for $\mathcal{D}$.*

The proof is in Appendix B.

## 4 Randomized vs. Truthful-in-Expectation Mechanisms

In this section, we show that for every two-bidder truthful-in-expectation mechanism $M$, there exists a universally truthful mechanism $M'$ with "the same" behavior. In other words, for two-bidder single item auctions, relaxing the solution concept from universal truthfulness to truthfulness in expectation is useless: every allocation function that is implementable by a truthful-in-expectation mechanism is also implementable by a universally truthful mechanism.

**Definition 4.1** *A mechanism $M'$ implements another mechanism $M$ if, for every bid vector and every bidder $i$, $M'$ allocates the item to $i$ with the same probability and at the same expected price as $M$.*

Notice that if $M'$ implements $M$, then the two mechanisms have the same expected revenue. We show that for two bidders, for every truthful-in-expectation mechanism $M$, there is a universally truthful mechanism that implements it. We would also like to show that this can be done *efficiently*. We assume that $M$ is represented by an oracle:

**Definition 4.2** *Let $\mathcal{R}_u^1$ be $\{(x,p) \mid \exists v \text{ such that } M(v,u)_1 = (x,p)\}$, and $\mathcal{R}_v^2$ be $\{(x,p) \mid \exists u \text{ such that } M(v,u)_2 = (x,p)\}$. An* alternative oracle *gets as input the value $v$ of bidder $i$ and returns $\mathcal{R}_v^{3-i}$.*

**Theorem 4.3** *If $M$ is a truthful-in-expectation ex-post IR mechanism for two bidders, then there exists a universally truthful ex-post IR mechanism $M'$ that implements $M$. Moreover, if $M$ is represented by an alternative oracle then $M'$ can be found in time that is polynomial in $\max_{i,v} |\mathcal{R}_v^i|$.*

The proof is constructive and presents an algorithm that takes a truthful-in-expectation mechanism $M$ as input and outputs a universally truthful mechanism $M'$ that implements $M$. The algorithm is efficient as long as the alternative oracle can be efficiently implemented. In particular we get that:

**Corollary 4.4** *There is an algorithm that finds a 2-bidder optimal revenue-maximizing deterministic truthful mechanism in time polynomial to the size of the support of the distribution.*

**Proof:** In the previous section we showed that an optimal truthful-in-expectation mechanism $M$ can be found in time polynomial in the size of the support of the distribution. Since alternative oracle queries can be answered in time polynomial in the size of the support of the mechanism, Theorem 4.3 guarantees that we can find a universally truthful mechanism $M'$ that implements $M$ in polynomial time. Finally, $M'$ is a distribution over deterministic mechanisms that achieves an optimal revenue (in expectation). Thus, every mechanism in the support of $M$ obtains that revenue, in particular the one that sets all the random coins to, say, 0. □

This amounts to a re-derivation of a result in [11]. (The statement of their result was made public before we started working on this paper.) However, we would like to stress that the applicability of our result is not limited to revenue maximization. For example, Vincent and Manelli show the following:

**Theorem 4.5 (Vincent and Manelli [7])** *Let $M$ be a Bayesian truthful mechanism with independent distributions. Then, there is a truthful-in-expectation mechanism that implements[7] $M$.*

Therefore, by [7] and Theorem 4.3, we have that:

**Theorem 4.6** *Let $M$ be a two-bidder Bayesian truthful mechanism with independent distributions. Then there is a universally truthful mechanism that implements $M$.*

Furthermore, we show the following result that together with the algorithm answers an open question of Mehta and Vazirani [8]. (For the proof see Appendix C.2 Theorem C.5 and Appendix D Lemma D.1.)

**Theorem 4.7** *Every truthful-in-expectation mechanism for n bidders where each bidder has only two possible values can be implemented as a universally truthful mechanism. However, there exists a truthful-in-expectation mechanism for three bidders where each bidder has three possible values that cannot be implemented as a universally truthful mechanism.*

Back to designing optimal auctions, we also show in Appendix D an explicit distribution with small support size, for which the best truthful-in-expectation mechanism performs better than the best deterministic mechanism.

**Proposition 4.8** *There exists a distribution $\mathcal{D}$ for three bidders, with each bidder having only four possible values in the support, for which the optimal truthful-in-expectation mechanism achieves strictly more revenue than any deterministic mechanism.*

### 4.1 The Algorithm

For the rest of the discussion in this subsection we fix a truthful in expectation mechanism $M$. In this presentation of the algorithm we assume $\mathcal{R}_v^i$ contains finitely many elements. We extend it in Appendix C.1 to a procedure for constructing a universally truthful mechanism that implements M, even when the sets $\mathcal{R}_v^i$ are infinite.

Observe that by truthfulness, if $(x^i, p^i) \in \mathcal{R}_{v_{-i}}^i$, then $(x^i, \hat{p}^i) \notin \mathcal{R}_{v_{-i}}^i$, for every $p \neq \hat{p}$. Hence we define a total order on the elements of $\mathcal{R}_{v_{-i}}^i$: $(x^i, p^i) > (\hat{x}^i, \hat{p}^i)$ if and only if $x^i > \hat{x}^i$. We can therefore sort the elements in $\mathcal{R}_{v_{-i}}$ in the increasing order, so that $(x_1^i, p_1^i) < (x_2^i, p_2^i) < \ldots < (x_\ell^i, p_\ell^i)$. If $x_\ell^i < 1$, then we add $(x_{\ell+1}^i, \infty)$ to $\mathcal{R}_{v_{-i}}^i$. If $(0, 0)$ is not in $\mathcal{R}_{v_{-i}}^i$, then add $(x_0^i = 0, p_0^i = 0)$ to $\mathcal{R}_{v_{-i}}^i$. Note that by individual rationality, $p_0^i$ must be 0.

The main idea of the algorithm is to set a take-it-or-leave-it offer to each bidder where the price of the offer is obtained by randomly "simulating" Myerson's payment formula [9]. The novelty of the proof is in showing that these offers can be coordinated in the following sense: when one bidder accepts the offer the other one will reject it. We present the algorithm as follows. The proof of its correctness is given in Appendix C.1. The running time of the algorithm is obviously polynomial in $\max_{i,v} |\mathcal{R}_v^i|$.

---

[7]A truthful-in-expectation mechanism $M'$ implements a Bayesian truthful mechanism $M$ (when the distribution is $\mathcal{D}$) if, for every each bidder $i$ and $v_i \in \mathcal{D}_i$, the probability with which bidder $i$ gets the item when he bids $v_i$ in $M$ is exactly the same as that in $M'$, i.e., $\sum_{v_{-i}:(v_i,v_{-i})\in\mathcal{D}} M'(v_i, v_{-i})_i$.

1. Sample $r \in [0,1]$ uniformly at random.

2. Elicit a bid from each bidder. Let the bids be $(v_1, v_2)$.

3. For each bidder $i$, $i \in \{1,2\}$:

   (a) Consider $\mathcal{R}^i_{v_{-i}} = \{(x^i_k, p^i_k) \mid x^i_k < x^i_{k+1}, \forall k\}$. Label points in each interval $(x^i_k, x^i_{k+1}] \subseteq [0,1]$ with $\frac{p^i_{k+1} - p^i_k}{x^i_{k+1} - x^i_k}$.
   
   (b) If $i = 1$, let $p$ be the label of the point $r$. If $i = 2$, let $p$ be the label of the point $1 - r$.
   
   (c) If $i = 1$, allocate the item to bidder $i$ only if $r \leq M(v_1, v_2)_i$. If $i = 2$, allocate the item to bidder $i$ only if $1 - r < M(v_1, v_2)_i$. If the bidder is allocated the item then his payment is $p$.

## 5 The Approximation Ratio of the $k$-Lookahead Auction

In this section we analyze the $k$-lookahead auction which is defined as follows. Find the $k$ bidders with the highest values, and denote this set of bidders by $K$. Run the revenue-maximizing truthful auction for $K$ conditioned on the values of bidders in $N \setminus K$. Notice that the auction for $K$ can either be the optimal truthful-in-expectation mechanism or the optimal deterministic mechanism.

**Theorem 5.1** *The approximation ratio of the $k$-lookahead auction is at least $\frac{3k-1}{2k-1}$. In particular, for $k = 2$ the approximation ratio is at least $\frac{5}{3}$, and the approximation ratio tends to $\frac{3}{2}$ as $k$ tends to $\infty$.*

We remark that if the auction for $K$ is allowed to be truthful-in-expectation, then the approximation ratio is with respect to the optimal truthful-in-expectation mechanism[8]. Taking into account the result of Section 3 we get that:

**Corollary 5.2** *The $k$-lookahead auction provides a $\frac{3k-1}{2k-1}$-approximation to the revenue of the optimal truthful-in-expectation mechanism. The auction runs in polynomial time in the oracle model.*

Our derandomization result for 2-bidder auctions (Section 4) gives us that:

**Corollary 5.3** *The 2-lookahead auction provides a $\frac{5}{3}$-approximation to the revenue of the optimal truthful-in-expectation mechanism. The 2-lookahead auction is deterministic in this case and runs in polynomial time in the oracle model.*

In particular this implies that we can bound the ratio between the revenue achieved by the optimal truthful-in-expectation mechanism and the optimal deterministic mechanism for $n$-bidder single-item auctions in general:

**Corollary 5.4** *Let $r$ be the revenue of the optimal truthful-in-expectation mechanism for a single item with distribution $\mathcal{D}$. There is a deterministic mechanism for a single item and distribution $\mathcal{D}$ with revenue at least $\frac{3r}{5}$.*

---
[8]Similarly, if the auction for $K$ is allowed to be Bayesian truthful, then the approximation ratio is with respect to the optimal Bayesian truthful mechanism.

## 5.1 Analysis of the $k$-Lookahead Auction

Denote the original distribution by $\mathcal{D}$, and denote the conditional distribution of the values of the bidders in $K$ given the values of bidders in $N \setminus K$ by $\mathcal{D}_K$. We let $v_{k+1}$ denote the the value of the $(k+1)^{\text{th}}$-highest bidder. We show that one of the following three families of auctions provides a good approximation ratio. The $k$-lookahead auction obviously provides at least as much expected revenue, and the theorem follows. The auctions are defined for $k \geq 2$. The second and third auctions depend on a parameter $t \geq 1$, to be specified later.

1. **$k$-Highest Auction:** Run the optimal auction. If one of the bidders in $N \setminus K$ is assigned the item in the optimal auction, no bidder is assigned the item and no one is charged anything. If one of the bidders in $K$ is assigned the item in the revenue-maximizing auction then assign him the item and charge him as in the revenue-maximizing auction.

2. **$t$-Fixed Price Auction:** Select one bidder ("the reserve bidder") from $K$ uniformly at random, denote this bidder by $i$. If any of the bidders in $K \setminus \{i\}$ has value above $t \cdot v_{k+1}$ then he receives the item and pays $t \cdot v_{k+1}$. If there are several such bidders, break ties arbitrarily. Otherwise, the reserve bidder gets the item and pays $v_{k+1}$.

3. **$t$-Pivot Auction:** Select one bidder ("the pivot") from $K$ uniformly at random, and denote this bidder by $i$. If any of the bidders of in $K \setminus \{i\}$ has value above $t \cdot v_{k+1}$ then run the revenue maximizing auction for bidders in $K$, conditioned on the values of bidders in $N \setminus K$. Otherwise the pivot bidder gets the item and pays $v_{k+1}$.

It is straightforward to see that the $k$-Highest Auction and the $t$-Fixed Price Auction are truthful and individually rational[9]. To see that the $t$-Pivot Auction is truthful we observe that this auction is monotone: the only non-straightforward case to check is when bidder $i$ raises his value and forces the mechanism to run the optimal auction. However, in this case bidder $i$ was not allocated the item before raising his value, so monotonicity is preserved.

**Proof:** (of Theorem 5.1) Let $l$ be the event where no bidder in $K$ has value at least $t \cdot v_{k+1}$, and let $\bar{l}$ be the complement of this event. We partition the expected revenue of the optimal auction:

- let $L_l$ be the expected revenue from bidders in $N \setminus K$ from instances where event $l$ occurs.
- let $L_{\bar{l}}$ be the expected revenue from bidders in $N \setminus K$ from instances where event $\bar{l}$ occurs.
- Let $M$ be the expected revenue from bidders in $K$ from instances where event $l$ occurs.
- Let $H$ be the expected revenue from bidders in $K$ from instances where event $\bar{l}$ occurs.

Observe that the expected revenue of the optimal auction is $L_l + L_{\bar{l}} + H + M$. We continue by proving several lemmas.

**Lemma 5.5** *The expected revenue of the $k$-Highest Auction is $M + H$.*

**Proof:** By definition the auction extracts exactly the same revenue as the optimal auction from bidders in $K$ and no revenue from bidders in $N \setminus K$. The lemma follows. □

**Lemma 5.6** *The expected revenue of the $t$-Fixed Price Auction is at least $L_l + \frac{M}{t} + \frac{k-1}{k} \cdot t \cdot L_{\bar{l}} + \frac{1}{k} \cdot L_{\bar{l}}$.*

---

[9]If the optimal auctions used by the $k$-Highest Auction and the $t$-Pivot Auction are deterministic or universally truthful, then all three auctions are universally truthful, as is their convex combination. In this case our proof shows that there is a deterministic auction on the $k$ highest bidders that achieves a $\left(\frac{3k-1}{2k-1}\right)$-approximation to the deterministic optimal auction.

**Proof:** First, notice that the revenue of the $t$-Fixed Price Auction in every instance is at least $v_{k+1}$ (either the reserve bidder is allocated the item and pays $v_{k+1}$ or the auction sells the item at a higher price). Suppose that event $l$ occurs. This case contributes $L_l + M$ to the expected revenue of the optimal auction. Observe that, if $l$ occurs, in any instance where the optimal auction sells the item to bidders in $N \setminus K$, its revenue is at most $v_{k+1}$ (the price for a sold item is at most the value of the bidder), and that in any instance the optimal auction sells the item to bidders in $N \setminus K$ the revenue is at most $t \cdot v_{k+1}$. Thus, the instances where event $l$ occurs contribute $L_l + \frac{M}{t}$ to the expected revenue of the $t$-Fixed Price Auction.

Suppose now that event $\bar{l}$ occurs. Thus, there exists some bidder $b$ with $v_b > t \cdot v_{k+1}$. With probability exactly $\frac{k-1}{k}$, $b$ is not the reserve bidder and in this case the revenue of the auction is $t \cdot v_{k+1}$. With probability $\frac{1}{k}$ we have that $b$ is the reserve bidder and the revenue of the auction is at least $v_{k+1}$. In particular, for every instance where the optimal auction sells the item to bidders in $N \setminus K$ (at a price of at most $v_{k+1}$) the $t$-Fixed Price Auction has an expected revenue of at least $\frac{k-1}{k} \cdot t \cdot L_{\bar{l}} + \frac{1}{k} \cdot L_{\bar{l}}$. Together with the contribution from instances where event $l$ occurs we have that the expected revenue of the auction is at least $L_l + \frac{M}{t} + \frac{k-1}{k} \cdot t \cdot L_{\bar{l}} + \frac{1}{k} \cdot L_{\bar{l}}$. □

**Lemma 5.7** *The expected revenue of the $t$-Pivot Auction, conditioned on the values of bidders in $K$, is at least $L_l + \frac{M}{t} + \frac{k-1}{k} H + \frac{1}{k} L_{\bar{l}}$.*

**Proof:** Suppose that event $l$ occurs. The revenue of the $t$-Pivot Auction in every instance where $l$ occurs is $v_{k+1}$. The expected revenue of the optimal auction in this case is $L_l + M$, and similarly to the the analysis of the $t$-Fixed Price Auction the expected contribution to the revenue from instances where event $l$ occurs is $L_l + \frac{M}{t}$.

Suppose that event $\bar{l}$ occurs. Thus, there exists some bidder $b$ with $v_b > t \cdot v_{k+1}$. With probability exactly $\frac{k-1}{k}$, $b$ is not the reserve bidder and in this case the revenue of the auction is at least $H$. With probability $\frac{1}{k}$ we have that $b$ is the reserve bidder and the revenue of the auction is at least $\frac{1}{k} \cdot v_{k+1}$. Again, similarly to the analysis of the $t$-Fixed Price Auction the expected contribution to the revenue when the event $\bar{l}$ occurs is $\frac{k-1}{k} H + \frac{1}{k} L_{\bar{l}}$. Overall, the expected revenue of the auction is at least $L_l + \frac{M}{t} + \frac{k-1}{k} H + \frac{1}{k} L_{\bar{l}}$. □

Next we need some definitions. Conditioned on the values of bidders in $N \setminus K$, let OPT be the revenue of the revenue-maximizing auction, $R_h$ be the revenue of the $k$-Highest Auction, $R_f$ be the expected revenue of the $t$-Fixed Price Auction, $R_p$ the expected revenue of the $t$-Pivot Auction and $R = \max(R_f, R_p)$. In addition, for the rest of the proof we fix $t = \frac{2k-1}{k-1}$.

**Lemma 5.8** $R \geq L_l + L_{\bar{l}} + \frac{H+M}{t}$.

**Proof:** We divide the analysis into two cases. Suppose first that $L_{\bar{l}} \cdot t \geq H$, which implies that $\frac{k-1}{k} \cdot (t-1) \cdot L_{\bar{l}} = L_{\bar{l}} \geq \frac{H}{t}$ by our choice of $t$. We have that:

$$R_f \geq L_l + \frac{M}{t} + \frac{k-1}{k} \cdot t \cdot L_{\bar{l}} + \frac{1}{k} \cdot L_{\bar{l}} \geq L_l + \frac{M}{t} + L_{\bar{l}} + \frac{k-1}{k} \cdot (t-1) \cdot L_{\bar{l}} \geq L_l + L_{\bar{l}} + \frac{H+M}{t}$$

Suppose now that $L_{\bar{l}} \cdot t < H$:

$$R_p \geq L_l + \frac{M}{t} + \frac{k-1}{k} \cdot H + \frac{1}{k} \cdot L_{\bar{l}} \geq L_l + \frac{M}{t} + \frac{H}{t} + \frac{k^2 - 2k + 1}{k(2k-1)} \cdot H + \frac{1}{k} \cdot L_{\bar{l}} > L_l + \frac{M}{t} + \frac{H}{t} + \frac{k-1}{k} \cdot L_{\bar{l}} + \frac{1}{k} \cdot L_{\bar{l}}$$

$$= L_l + L_{\bar{l}} + \frac{H+M}{t}$$

□

We are now finally able to analyze the ratio between the expected ratio of the revenue-maximizing auction and the $k$-lookahead auction. We consider two cases and show that in each one the expected ratio is at most $2 - \frac{1}{t} = \frac{3k-1}{2k-1}$. In the first case we assume that $L_l + L_{\bar{l}} \leq (H+M)(1 - \frac{1}{t})$. Therefore,

$$\frac{OPT}{R_h} \leq \frac{L_l + L_{\bar{l}} + H + M}{H + M} \leq \frac{(H+M)(1 - \frac{1}{t}) + H + M}{H + M} = 2 - \frac{1}{t}$$

Now assume that $L_l + L_{\bar{l}} > (H+M)(1 - \frac{1}{t})$. We have that, using Lemma 5.8:

$$\frac{OPT}{R} \leq \frac{L_l + L_{\bar{l}} + H + M}{L_l + L_{\bar{l}} + \frac{H+M}{t}} < \frac{(H+M)(1 - \frac{1}{t}) + H + M}{(H+M)(1 - \frac{1}{t}) + \frac{H+M}{t}} = 2 - \frac{1}{t}.$$

□

# References

[1] Mark Armstrong. Optimal multi-object auctions. *Review of Economic Studies*, 67(3):455–81, July 2000.

[2] Sayan Bhattacharya, Gagan Goel, Sreenivas Gollapudi, and Kamesh Munagala. Budget constrained auctions with heterogeneous items. In *STOC*, pages 379–388, 2010.

[3] Shuchi Chawla, Jason D. Hartline, David L. Malec, and Balasubramanian Sivan. Multi-parameter mechanism design and sequential posted pricing. In *STOC*, pages 311–320, 2010.

[4] Jacques Cremer and Richard P McLean. Optimal selling strategies under uncertainty for a discriminating monopolist when demands are interdependent. *Econometrica*, 53(2):345–61, March 1985.

[5] Paul Klemperer. Auction theory: A guide to the literature. Technical report, EconWPA, March 1999.

[6] Benny Lehmann, Daniel Lehmann, and Noam Nisan. Combinatorial auctions with decreasing marginal utilities. In EC'01.

[7] Alejandro M. Manelli, Daniel, and R. Vincent. Bayesian and dominant strategy implementation in the interdependent private values model. Forthcoming in Econometrica.

[8] Aranyak Mehta and Vijay V. Vazirani. Randomized truthful auctions of digital goods are randomizations over truthful auctions. In *EC'04*.

[9] R. B. Myerson. Optimal auction design. *Mathematics of Operations Research*, 6(1):58–73, 1981.

[10] Noam Nisan. 2007. Introduction to Mechanism Design (for Computer Scientists). In "Algorithmic Game Theory", N. Nisan, T. Roughgarden, E. Tardos and V. Vazirani, editors.

[11] Christos Papadimitiou and George Pierrakos. Private communication.

[12] Amir Ronen. On approximating optimal auctions. In *ACM Conference on Electronic Commerce*, pages 11–17, 2001.

[13] Amir Ronen and Amin Saberi. On the hardness of optimal auctions. In *FOCS*, pages 396–405, 2002.

# A  Mechanisms for Multi-Parameter Domains

In this section we extend our technique of Section 3 to several multi-parameter settings and obtain optimal auctions in these settings. We first set up a general framework that relates a linear program to revenue maximizing mechanism design, then we explore specific settings with this perspective.

Any truthful-in-expectation mechanism $\mathcal{M}$ can be described by the following parameters: at a bid vector $\mathbf{v}$, $\mathcal{M}$ allocates to bidder $i$ a fraction $x_{i,S,\mathbf{v}}$ of each bundle $S \subseteq M$, and charges him an expected price $p_{i,\mathbf{v}}$. We want $\mathcal{M}$ to be feasible, individually rational and truthful-in-expectation, and maximize the revenue. We use the following LP:

*Maximize:* $\sum_{\mathbf{v} \in \mathcal{D}} \Pr_{\mathcal{D}}(\mathbf{v}) \cdot p_{i,\mathbf{v}}$, s.t.

- $\forall i, \mathbf{v} \in \mathcal{T}_i, S \subseteq M : \sum_i x_{i,S,\mathbf{v}} \leq 1$
- $\forall j \in M, \mathbf{v} \in \mathcal{T}: \sum_{i, j \in S} x_{i,S,\mathbf{v}} \leq 1$
- $\forall i, (\mathbf{v}_i, \mathbf{v}_{-i}) \in \mathcal{T}_i, (\mathbf{v}'_i, \mathbf{v}_{-i}) \in \mathcal{T}_i$:

$$\sum_{S \subseteq M} v_i(S) x_{i,S,(\mathbf{v}_i, \mathbf{v}_{-i})} - p_{i,(\mathbf{v}_i, \mathbf{v}_{-i})} \geq \sum_{S \subseteq M} v_i(S) x_{i,S,(\mathbf{v}'_i, \mathbf{v}_{-i})} - p_{i,(\mathbf{v}'_i, \mathbf{v}_{-i})}$$

- $\forall i, \forall S \subseteq M, \mathbf{v} \in \mathcal{T}: x_{i,S,\mathbf{v}} \geq 0, p_{i,\mathbf{v}} \geq 0$
- $\forall i, \mathbf{v}_{-i} \in \mathcal{D}_{-i} : p_{i,(\mathbf{0},\mathbf{v}_{-i})} = 0$.

**Definition A.1** *Given a solution $\{x_{i,S,\mathbf{v}}, p_{i,\mathbf{v}}\}_{i \in N, S \subseteq M, \mathbf{v} \in \mathcal{T}}$ to the LP, we say that a mechanism $\mathcal{M}$ $\alpha$-decomposes the solution if, at each bid vector $\mathbf{v} \in \mathcal{T}_i$, for each bundle $S$ and bidder $i$, $\mathcal{M}$ allocates exactly $x_{i,S,\mathbf{v}}$ fraction of bundle $S$ to bidder $i$, and charges the bidder an expected price of $p_{i,\mathbf{v}}$. When $\alpha$ is $1$, we say that $\mathcal{M}$ decomposes the solution.*

**Observation A.2** *If a mechanism $\mathcal{M}$ $\alpha$-decomposes an optimal solution to the LP above, then $\mathcal{M}$ is truthful-in-expectation and the revenue of $\mathcal{M}$ is at least an $\alpha$-approximation to the optimal auction.*

Proof of the observation is essentially the same as that of Theorem 3.1. The problem of designing optimal mechanism, however, cannot be directly reduced to solving the LP. First, the number of variables in the LP is exponential in $m$, and each truthfulness constraint involves exponentially many variables. Second, even if we can solve the LP optimally, not all solutions can be decomposed by mechanisms. An easy counterexample is when $m = n = 3$, where each bidder $i$ single-mindedly demands the bundle $M \setminus \{i\}$, having a valuation of 1 for it. Let $\mathbf{v}$ be the only bid vector in this case, then the solution where $x_{i, M \setminus \{i\}, \mathbf{v}} = \frac{1}{2}$, $p_{i,\mathbf{v}} = \frac{1}{2}$ for all $i$ is an optimal solution to the LP, but it is obvious that no mechanism can decompose this solution.

In the following subsections, we show that we can overcome both obstacles for the settings of additive valuations and unit-demand valuations.

## A.1  Additive Valuations

In the additive domain, each bidder's valuation $v_i$ is determined by his valuation of each item in $M$, and $v_i(S)$ for any $S \subseteq M$ is simply $\sum_{j \in S} v_i(j)$. If we use $x_{i,j,\mathbf{v}}$ to denote the fraction of item $j$ being assigned to bidder $i$ at bid vector $\mathbf{v}$, then

$$\sum_{S \subseteq M} x_{i,S,\mathbf{v}} v_i(S) = \sum_{S \subseteq M} x_{i,S,\mathbf{v}} \sum_{j \in S} v_i(j) = \sum_{j \in M} v_i(j) \sum_{S \ni j} x_{i,S,\mathbf{v}} = \sum_{j \in M} v_i(j) x_{i,j,\mathbf{v}}.$$

Therefore we are able to express the truthful-in-expectation condition in the LP in terms of polynomially many variables $\{x_{i,j,\mathbf{v}}, p_{i,\mathbf{v}}\}_{i\in N, j\in M, \mathbf{v}\in\mathcal{T}}$. The feasibility constraints can also be easily expressed as

$$\forall \mathbf{v} \in \mathcal{T}, \forall j \in M, \sum_i x_{i,j,\mathbf{v}} \leq 1.$$

Additionally, we have that

$$\forall i, \forall j \in M, \forall \mathbf{v} \in \mathcal{T}, x_{i,j,\mathbf{v}} \geq 0, p_{i,\mathbf{v}} \geq 0;$$

$$\forall i, \forall \mathbf{v}_{-i} \in \mathcal{D}_{-i}, p_{i,(\mathbf{0},\mathbf{v}_{-i})} = 0,$$

then we have obtained a much smaller LP whose optimal solution has the same value as that of the original LP. Fortunately, we are able to decompose a solution for the new LP by a mechanism: given any feasible solution $\{x_{i,j,\mathbf{v}}, p_{i,\mathbf{v}}\}_{i\in N, j\in M, \mathbf{v}\in\mathcal{T}}$ for the LP, at bid vector $\mathbf{v}$,

- if $\mathbf{v}$ is in $\mathcal{T}$, for bidder $i$, if $\mathbf{v}_{-i}$ is not in $\mathcal{D}_{-i}$, allocate nothing to $i$; otherwise allocate each item $j \in M$ with probability $x_{i,j,\mathbf{v}}$ to bidder $i$. If bidder $i$ gets item $j$, then charge him a price $p_{i,\mathbf{v}} \cdot \frac{v_i(j)}{\sum_{j'\in M} v_i(j) x_{i,j',\mathbf{v}}}$, otherwise charge him nothing for this item.
- if $\mathbf{v}$ is not in $\mathcal{T}$, no bidder gets allocated anything, and no payment is made either.

It is obvious that the mechanism allocates the items according to $x_{i,j,\mathbf{v}}$'s. To see that it also charges the right expected prices, just notice that at any bid vector $\mathbf{v} \in \mathcal{T}_i$, the expected price for bidder $i$ is

$$\sum_{j\in M} x_{i,j,\mathbf{v}} p_{i,\mathbf{v}} \cdot \frac{v_i(j)}{\sum_{j'\in M} v_i(j') x_{i,j',\mathbf{v}}} = p_{i,\mathbf{v}}.$$

Since we can solve the new LP efficiently, we have proved the following theorem:

**Theorem A.3** *For additive valuations, we can compute an optimal truthful-in-expectation mechanism in time polynomial to the number of bidders, the number of items, and the size of the support of the valuation distribution.*

## A.2 Unit Demand Valuations

In this domain, each bidder $i$ has a valuation $v_i(j)$ for each item $j \in M$, and his valuation of a bundle $S \subseteq M$ is $\max_{j\in M} v_i(j)$. Therefore, each $v_i(\cdot)$ can also be described by a vector $\mathbf{v}_i \in \mathbb{R}^m$. For this domain we prove the following theorem:

**Theorem A.4** *For unit demand valuations, we can compute an optimal truthful-in-expectation mechanism in time polynomial to the number of bidders, the number of items, and the size of the support of the valuation distribution.*

Since each bidder does not distinguish a bundle $S$ of size larger than 1 and the item in $S$ that he values most, it is without loss of generality to assume that a mechanism allocates only bundles of size at most 1. Therefore in the LP, we have only variables $x_{i,j,\mathbf{v}}$ for all $i \in N$, $j \in M$ and $\mathbf{v} \in \mathcal{T}$, which allows us to solve the LP efficiently. Therefore it remains only to design a mechanism that implements a solution to the LP.

We observe that every solution to the LP can be described by a set of matrices $\{X^{\mathbf{v}}\}_{\mathbf{v}\in\mathcal{T}}$, where for each $\mathbf{v}$, $X^{\mathbf{v}}$ is a $n \times m$ matrix whose entries are nonnegative and satisfy the following constraints: (i) for each row $i \in N$, $\sum_{j\in M} X_{i,j} \leq 1$; (ii) for each column $j \in M$, $\sum_{i\in M} X_{i,j} \leq 1$. To implement a solution means that for every $\mathbf{v} \in \mathcal{D}$, the mechanism allocates item $j$ to bidder $i$ with probability $X_{i,j}^{\mathbf{v}}$, and in each realization, no bidder should get more than one item; in other words, each realization of the mechanism should be a $n \times m$ matrix whose entries are from $\{0, 1\}$, subject to the same constraints (i)(ii). The problem therefore boils down to expressing a matrix in $\mathbb{R}^{n \times m}$ satisfying (i)(ii) as a convex combination of matrices in $\{0, 1\}^{n \times m}$.

When $m$ equals $n$, and if all constraints (i)(ii) hold with equality, then the problem is exactly expressing a doubly stochastic matrix by a convex combination of permutation matrices. This is exactly what the classical Birkhoff-von Neumann theorem tells us:

**Theorem A.5 (Birkhoff-von Neumann)** *Given any $n \times n$ doubly stochastic matrix, we can write it as a convex combination of permutation matrices, and the coefficients can be computed in time polynomial in $n$.*

Therefore, if we can extend a general $n \times m$ matrices that satisfy constraints (i)(ii) to some larger doubly stochastic matrix in polynomial time, then we would be able to implement any LP solution. This is exactly what we are going to do, which completes the proof of Theorem A.4.

**Lemma A.6** *Given any $n \times m$ matrix $A$ whose entries are nonnegative real numbers that satisfy*

$$\forall i \in [n], \sum_j A_{ij} \leq 1, \qquad \forall j \in [m], \sum_i A_{ij} \leq 1,$$

*we can output in polynomial time a square matrix $A'$ that is doubly stochastic, and the submatrix consisting of its first $n$ rows and $m$ columns is identical with $A$.*

**Proof:** Without loss of generality, suppose $n \leq m$. We first make it an $m \times m$ square matrix by appending $m - n$ all zero rows. Abusing the notation, we still call it $A$, as it still satisfies the condition in the theorem. Then we expand the matrix to a $2m \times 2m$ matrix $B$, which has four $m \times m$ submatrices. Therefore $B = \begin{bmatrix} A_1 & A_2 \\ A_3 & A_4 \end{bmatrix}$. We specify the four submatrices and show that $B$ is doubly stochastic. We let $A_1$ be $A$ itself, and $A_4$ be $A^T$, the transpose of $A$. Let $A_2$ be a diagonal matrix, whose $i$-th element on the diagonal is $1 - \sum_{j=1}^m A_{i,j}$; similarly, let $A_3$ be a diagonal matrix, whose $j$-th element on the diagonal is $1 - \sum_{i=1}^m A_{i,j}$. It is then obvious that for the first $m$ rows and first $m$ columns of $B$, the sum of elements in each of them is 1. For any of the last $m$ rows, the sum of the elements is $1 - \sum_{j=1}^m A_{i,j} + \sum_{j=1}^m A_{i,j} = 1$. The same argument shows that the sum of elements in each of the last $m$ columns is also 1. This completes the proof of the lemma. □

## A.3 OXS Valuations

In this domain, each bidder's valuation $v_i$ can be described by a bipartite graph $(V, E)$ with non-negative edge weights. The vertex set $V$ is partitioned into two sets: the set $M$ of items, and a set $C$ whose elements are called *clauses*. Edges go from items to clauses. For a bundle $B \subseteq M$, the valuation $v_i(B)$ is defined to be the sum of edge weights in a maximum-weight matching of the induced subgraph on $B \cup C$. This class includes both additive valuations (the special case that the graph is a matching from $M$ to $C$) and unit-demand valuations (the special case that the set $C$ consists of a single element). We can interpret each clause in $C$ as a demand that the bidder wishes to satisfy using a single element of $M$. The value of satisfying a clause depends on the element that is used to satisfy it, and the value of satisfying a set of clauses is the combined value of satisfying each one of them. We then interpret the value of a bundle to be the maximum value that can be achieved by satisfying a subset of clauses using distinct elements of the bundle.

**Definition A.7** *A type distribution is* simple *if its support consists of only two types — denoted by $v_L, v_H$ — that satisfy $v_L(B) \leq v_H(B)$ for every bundle $B$.*

**Lemma A.8** *Consider a simple type distribution with $\Pr(v_L) = \pi_L$ and $\Pr(v_H) = \pi_H = 1 - \pi_L$. Fix two distributions $D_L, D_H$ over bundles. There exists a truthful-in-expectation mechanism mapping $t_i$ to $D_i$ for $i \in \{L, H\}$, if and only if $v_L(D_L) + v_H(D_H) \geq v_L(D_H) + v_H(D_L)$. If the set of such mechanisms is nonempty, then the optimal revenue of any such mechanism (subject to individual rationality) is given by the formula*
$$\mathrm{OptRev}(D_L, D_H) = v_L(D_L) - \pi_H v_H(D_L) + \pi_H v_H(D_H).$$

**Proof:** If $v_L(D_L) + v_H(D_H) < v_L(D_H) + v_H(D_L)$ then any mechanism mapping $t_i$ to $D_i$ for $i \in \{L, H\}$ violates weak monotonicity and is consequently not truthful in expectation. Otherwise, consider the posted-price mechanism with prices
$$p_L = v_L(D_L)$$
$$p_H = v_H(D_H) - v_H(D_L) + v_L(D_L).$$

At these prices, type $v_L$ gets zero utility from receiving $D_L$ and paying $p_L$, but the utility from receiving $D_H$ and paying $p_H$ is non-positive by weak monotonicity. Hence truthful bidding is weakly dominant for $v_L$. Type $v_H$ gets utility $v_H(D_L) - v_L(D_L)$ regardless of whether her bid is $v_H$ or $v_L$; this is non-negative by our assumption of a simple type distribution. Hence our posted-price mechanism satisfies individual rationality and incentive compatibility.

The revenue of our mechanism is
$$(1 - \pi_H)v_L(D_L) + \pi_H(v_H(D_H) - v_H(D_L) + v_L(D_L)) = v_L(D_L) - \pi_H v_H(D_L) + \pi_H v_H(D_H).$$

To see that this is optimal, consider any other mechanism with allocation rule $t_i \mapsto D_i$ and prices $p'_L, p'_H$. Individual rationality implies $p'_L \leq p_L$. Incentive compatibility implies $p'_H \leq p'_L + v_H(D_H) - v_H(D_L)$, and the right side is less than or equal to $p_H$ because $p'_L \leq p_L = v_L(D_L)$. □

**Lemma A.9** *When the type distribution is simple, an optimal mechanism must give $v_H$ its utility maximizing bundle and must give $v_L$ the bundle $B$ that maximizes $v_L(B) - \pi_H v_H(B)$. In particular, for simple type distributions there is always an optimal mechanism that is deterministic.*

**Proof:** Immediate from the formula for $\mathrm{OptRev}(D_L, D_H)$. □

**Theorem A.10** *Let $\mathcal{V}$ be a class of valuation functions that is closed under scalar multiplication and such that the problem of maximizing $f(B) - g(B)$ ($f, g \in \mathcal{V}$) is NP-hard, even when restricted to instances such that $\max_{B \neq \emptyset}\{f(B)/g(B)\}$ is between $1$ and $2^{\mathrm{poly}(n)}$. Then it is NP-hard to design optimal truthful-in-expectation mechanisms for one bidder with a simple type distribution over valuations in $\mathcal{V}$.*

**Proof:** Given $f, g \in \mathcal{V}$ let $\lambda = \max_{B \neq \emptyset}\{f(B)/g(B)\}$. Assume $1 \leq \lambda \leq 2^{\mathrm{poly}(n)}$. Define types $v_L, v_H$ such that $v_L(B) = f(B)$ and $v_H(B) = \lambda g(B)$ for all $B$. Set $\pi_H = \lambda^{-1}$ and $\pi_L = 1 - \pi_H$. Our definition of $\lambda$ ensures that this type distribution is simple. We have seen in Lemma A.9 that constructing an optimal truthful-in-expectation mechanism is equivalent to finding the bundles $B_L, B_H$ that maximize
$$v_H(B_H) = \lambda g(B_H)$$
and
$$v_L(B_L) - \pi_H v_H(B_L) = f(B_L) - g(B_L).$$

In particular, constructing an optimal truthful-in-expectation mechanism requires maximizing $f(B_L) - g(B_L)$, which by hypothesis is NP-hard. □

**Theorem A.11** *It is NP-hard to design an optimal truthful-in-expectation mechanism for OXS bidders, even when there is a single bidder with a simple type distribution.*

**Proof:** By Theorem A.10, it suffices to prove that the problem of maximizing $f(B) - g(B)$ when $f, g$ belong to OXS is NP-hard. To do so, we reduce from Clique. Suppose we are given an undirected graph $G = (V, E)$ with $n$ vertices and $m$ edges, and we wish to decide if it contains a $k$-clique. Let the items be edges of $G$ and let the valuation functions be specified as follows.

$$f(B) = \min\left\{|B|, \binom{k}{2}\right\}$$

$$g(B) = \max\{|A| \;:\; A \subseteq B, \text{every connected component of } A \text{ is either acyclic or unicyclic}\}.$$

(Recall that a unicyclic graph is one that has exactly one simple cycle. Equivalently, it is a connected graph with $p$ vertices and $p$ edges, for some $p$.) Both of these are OXS valuation functions. Valuation $f$ is represented by $\binom{k}{2}$ clauses, each of which assigns value 1 to every edge of $G$. Valuation $g$ is represented by $n$ clauses, each of which is associated to a vertex $v$ of $G$ and assigns value 1 to every edge incident to $v$.

If $G$ contains a $k$-clique then the edges of this $k$-clique form a bundle such that $f(B) - g(B) = \binom{k}{2} - k = \binom{k-1}{2}$. Conversely, if $G$ contains an edge set $B$ such that $f(B) - g(B) \geq \binom{k-1}{2}$ then we may assume without loss of generality that $|B| \leq \binom{k}{2}$. If the graph $H = (V, B)$ has any connected component that is a tree, then we can remove the edges of this tree from $B$ without changing the value of $f(B) - g(B)$. Consequently, we can assume without loss of generality that every connected component of $H$ other than isolated vertices contains a unicyclic subgraph. Then $g(B)$ is the number of non-isolated vertices in $H$. The relation $g(B) = f(B) - (f(B) - g(B)) \leq \binom{k}{2} - \binom{k-1}{2} = k$ implies that $H$ has at most $k$ non-isolated vertices. Let $p$ be the number of such vertices, and let $H_0$ denote the induced subgraph on these $p$ vertices. The graph $H_0$ has $p \leq k$ vertices and $p + \binom{k-1}{2}$ edges, which can only happen if $p = k$ and $H_0$ is a $k$-clique. $\square$

## A.4 Single Parameter Domains

In this section, we show that in single-dimensional settings the problem of designing revenue maximizing truthful-in-expectation mechanisms can be reduced to finding an outcome that maximizes the social welfare.

**Definition A.12** *In a single-dimensional setting, each bidder $i$'s type is represented by a real number $v_i$, and for each outcome $\omega \in \Omega$, there is a constant $\alpha_{i,\omega}$ such that bidder $i$'s valuation for $\omega$ is $v_i \cdot \alpha_{i,\omega}$.*

For example, for combinatorial auctions with single-minded bidders with known bundles, each bidder has a publicly known bundle $S_i \subseteq M$, where $M$ is the set of items being sold. Each outcome $\omega$ is a partition $(T_1, T_2, \cdots, T_n)$ of $M$. Then $\alpha_{i,\omega}$ is 1 if and only if $T_i \supseteq S_i$, and 0 otherwise. Then each bidder's type is expressed by $v_i$, his valuation of any bundle that includes $S_i$.

**Theorem A.13** *In a single-dimensional setting, if there is an oracle that solves the social welfare maximization problem, then one can compute a revenue maximizing, truthful-in-expectation, ex-post IR mechanism in the explicit model.*

**Proof:** In a single-dimensional setting, the type of bidders can be represented by a vector $\mathbf{v}$, where $\mathbf{v}_i$ represents bidder $i$'s type. Let $\mathcal{D}$ be the support of the type distribution, and denote by $\Pr_{\mathcal{D}}(\mathbf{v})$ the probability of $\mathbf{v}$ occurring. Let $\mathcal{D}_{-i}$ be $\{\mathbf{v}_{-i} \mid \mathbf{v} \in \mathcal{D}\}$, $\mathcal{T}_i$ be $\{v_i \mid \exists v_{-i} \text{ s.t. } (v_i, \mathbf{v}_{-i}) \in \mathcal{D}\}$, and $\mathcal{T}$ be $\cup_i \mathcal{T}_i$.

For any randomized mechanism, given a bidding vector $\mathbf{v}$, it will output outcome $\omega$ with probability $x_{\mathbf{v},\omega}$, and charge bidder $i$ an expected price $p_{i,\mathbf{v}}$.

By a similar argument as that in Section 3, we can solve the optimal mechanism design problem if we could solve the following linear program:

$$\max \quad \sum_{\mathbf{v}\in\mathcal{D}} \Pr_{\mathcal{D}}(\mathbf{v}) \sum_i p_{i,\mathbf{v}}$$

$$\begin{array}{rcll}
\sum_{\omega\in\Omega} x_{\mathbf{v},\omega} & \leq & 1 & \forall \mathbf{v}\in\mathcal{D} \\
\mathbf{v}_i \sum_{\omega\in\Omega} \alpha_{i,\omega} x_{\mathbf{v},\omega} - p_{i,\mathbf{v}} & \geq & \mathbf{v}_i \sum_{\omega\in\Omega} \alpha_{i,\omega} x_{(v'_i,\mathbf{v}_{-i}),\omega} - p_{i,(v'_i,\mathbf{v}_{-i})}, & \forall i, \forall \mathbf{v}\in\mathcal{D}, \forall v'_i\in\mathcal{T}_i \\
p_{i,(0,\mathbf{v}_{-i})} & = & 0, & \forall i, \forall \mathbf{v}_{-i}\in\mathcal{D}_{-i} \\
x_{\mathbf{v},\omega} & \geq & 0, & \forall \omega\in\Omega, \forall \mathbf{v}\in\mathcal{D} \\
p_{i,\mathbf{v}} & \geq & 0, & \forall i, \forall \mathbf{v}\in\mathcal{D}
\end{array}$$

If we can optimally solve the LP, we can run a mechanism that outputs $\omega^*$ at a bid vector $\mathbf{v}$ with probability $x_{\mathbf{v},\omega^*}$, and charges bidder $i$ a payment of $p_{i,\mathbf{v}} \cdot \frac{x_{\mathbf{v},\omega^*}\alpha_{i,\omega^*}}{\sum_{\omega\in\Omega} x_{\mathbf{v},\omega}\alpha_{i,\omega}}$. As we will show below, the LP can be solved in polynomial time using the ellipsoid method, therefore the sum in the denominator has only polynomially many nonzero entries, and hence can be computed efficiently.

Since $\Omega$ can be exponentially large, the LP above may involve exponentially many variables, but it has only polynomially many constraints. Writing out its dual, we get

$$\min \gamma$$

$$\begin{array}{rcll}
\gamma + \sum_i \mathbf{v}_i \alpha_{i,\omega}(y_{i,\mathbf{v}} - \sum_{v'_i\in\mathcal{T}_i\setminus\{v_i\}} y_{i,(v'_i,\mathbf{v}_{-i})}) & \geq & 0, & \forall \mathbf{v}\in\mathcal{T}, \forall \omega\in\Omega \\
z_{i,\mathbf{v}} - \sum_{v'_i\in\mathcal{T}_i\setminus\{\mathbf{v}_i\}} z_{i,(v'_i,\mathbf{v}_{-i})} & \geq & \Pr_{\mathcal{D}}(\mathbf{v}), & \forall i, \forall \mathbf{v}\in\mathcal{T}, \text{ s.t. } \mathbf{v}_i\neq 0 \\
y_{i,\mathbf{v}} & \geq & 0, & \forall i, \forall \mathbf{v}\in\mathcal{D} \\
z_{i,\mathbf{v}} & \geq & 0, & \forall i, \forall \mathbf{v}\in\mathcal{D} \\
\gamma & \geq & 0
\end{array}$$

The dual has polynomially many variables but exponentially many constraints. As long as there is a separation oracle, we can solve the dual. Since there are only polynomially many constraints in the second group, it suffices to give a separation oracle for the first group. Since there are only polynomially many $\mathbf{v}\in\mathcal{D}$, this boils to the problem where given any $\mathbf{v}\in\mathcal{D}$ and $\gamma$, we are asked whether there is an $\omega\in\Omega$ such that

$$\sum_i \mathbf{v}_i \left( \sum_{v'_i\in\mathcal{T}_i\setminus\{v_i\}} y_{i,(v'_i,\mathbf{v}_{-i})} - y_{i,\mathbf{v}} \right) \alpha_{i,\omega} > \gamma.$$

This problem, however, can be solved by an oracle that, given a bid vector, outputs an outcome that maximizes the social welfare — feed the oracle with a bidding vector where bidder $i$'s type is $\mathbf{v}_i \left( \sum_{v'_i\in\mathcal{T}_i\setminus\{v_i\}} y_{i,(v'_i,\mathbf{v}_{-i})} - y_{i,\mathbf{v}} \right)$, and the left hand side becomes the social welfare of outcome $\omega$. If the $\omega$ that the oracle returns has social welfare at most $\gamma$, then no constraint for this $\mathbf{v}$ is violated, otherwise we find a violated constraint. This completes the proof. □

We remark that the social welfare maximization oracle in the theorem should be powerful enough that even when bidders' valuations are negative, it should still be able to output the best outcome. This requirement is no stricter than that in Myerson's original reduction for single item auctions, as in that case one may have negative virtual valuations.

## B  Appendix for Section 3

### B.1  Proof of Theorem 3.1

The proof consists of the following lemmas.

**Lemma B.1** *The algorithm is feasible, individually rational and truthful in expectation.*

**Proof:** If $\vec{v} \in \mathcal{T}$ then feasibility is ensured by the conditions of the LP. If $\vec{v} \notin \mathcal{T}$ then at most one bidder is allocated the item (with probability smaller than 1) and hence the algorithm is truthful.

To see that the algorithm is individually rational, assume that bidder $i$ wins the item and that his payment is $\frac{p_{i,\vec{v}}}{x_{i,\vec{v}}}$. By the constraints of the LP we have that $x_{i,\vec{v}} \cdot v_i - p_{i,\vec{v}} \geq 0$, since $(0, \vec{v}_{-i}) \in \mathcal{D}$. Hence $v_i \geq \frac{p_{i,\vec{v}}}{x_{i,\vec{v}}}$ as needed.

We now show that the algorithm is truthful in expectation. Consider bidder $i$. If $v_{-i} \notin \mathcal{D}_{-i}$ bidder $i$ never receives the item regardless of his value $v_i$. Hence assume that $v_{-i} \in \mathcal{D}_i$. Bidder $i$ wins the item with probability $x_{i,\vec{v}}$ and his payment in this case is $\frac{p_{i,\vec{v}}}{x_{i,\vec{v}}}$. Thus the expected profit of bidder $i$ when he declares his true value is $x_{i,\vec{v}} \cdot v_i - p_{i,\vec{v}}$. We now show that the profit of $i$ is not bigger when he declares any other value $v'_i$. We assume that $v \in \mathcal{T}$: otherwise by the definition of the algorithm bidder $i$ is facing several alternatives that their price does not depend on $i$'s value and taking the most profitable one, and truthfulness is obvious in that case.

Suppose that $(v'_i, v_{-i}) \in \mathcal{T}$. In this case, similarly to before, the profit of bidder $i$ is $x_{i,(v'_i,v_{-i})} \cdot v_i - p_{i,(v'_i,v_{-i})}$. By the constraints of the LP we have that $x_{i,\vec{v}} \cdot v_i - p_{i,\vec{v}} \geq x_{i,(v'_i,v_{-i})} \cdot v_i - p_{i,(v'_i,v_{-i})}$, as needed.

Now suppose that $(v'_i, v_{-i}) \notin \mathcal{T}$. Let $w = \arg\max_{v'_i \in V_i} x_{i,(v'_i,v_{-i})} \cdot v'_i - p_{i,(v'_i,v_{-i})}$. The profit of $i$ when declaring $v'_i$ is $x_{i,(w,v_{-i})} \cdot v_i - p_{i,(w,v_{-i})}$ which is at most, by the constraints of the LP, his profit from declaring his true value $v_i$ (i.e., $x_{i,\vec{v}} \cdot v_i - p_{i,\vec{v}}$). □

The following lemma is straightforward, since the size of the LP is polynomial in the size of $\mathcal{T}$, which is polynomial in the size of $\mathcal{D}$:

**Lemma B.2** *The algorithm runs in time polynomial in the size of $\mathcal{D}$.*

**Lemma B.3** *The expected revenue of the algorithm (over $\mathcal{D}$) equals the optimal value of the linear program. In addition, the optimal value of the linear program equals the revenue of the optimal (individually rational) truthful in expectation mechanism.*

**Proof:** Given that the valuation profile is $\vec{v} \in \mathcal{D}$, the expected payment from each bidder $i$ is $x_{i,\vec{v}} \cdot \frac{p_{i,\vec{v}}}{x_{i,\vec{v}}} = p_{i,\vec{v}}$. Thus the expected revenue of the algorithm (where the expectation is taken over $\mathcal{D}$) is exactly the optimal value of the linear program: $\sum_{\vec{v} \in \mathcal{D}} \Pr_\mathcal{D}[\vec{v}] \sum_i p_{\vec{v}}$.

We now prove the second statement. Consider the revenue maximizing, individually rational, truthful in expectation mechanism for $\mathcal{D}$. Set the variables of the linear program: for each $\vec{v} \in \mathcal{T}$, set $x_{i,\vec{v}}$ be the probability that $i$ receives the item given that the valuation profile is $\vec{v}$, and let $p_{i,\vec{v}}$ be his expected payment in case the valuation profile is $\vec{v}$. Notice that the linear program is feasible (the item is allocated to only one bidder so $\sum_i x_{i,\vec{v}} \leq 1$) and that the expected revenue equals the objective function of the linear program. □

# C Appendix for Section 4

## C.1 Correctness of the Algorithm in Section 4.1

The proof consists of the following claims.

**Claim C.1** *For each bidder $i$, the labels of the interval $[0, 1]$ are increasing. Moreover, suppose that bidder $i$ with valuation $v_i$ is assigned a fraction $x$ of the item in $M$, then all labels in $[0, x]$ are at most $v_i$, and all labels in $(x, 1]$ are at least $v_i$.*

**Proof:** By definition of $\mathcal{R}^i_{v_{-i}}$, $x$ is equal to $x^i_k$ for some $k$. The expected price for bidder $i$ in $M$ is therefore $p^i_k$. Let $v'_i$ be such that $M(v'_i, v_{-i}) = x^i_{k-1}$, and let the expected price in this case be $p'^i_k$. Let $v''_i$ be such that $M(v''_i, v_{-i}) = x^i_{k+1}$, and the expected price in this case be $p''^i_k$. From the definition of truthfulness, we have:

$$v_i \cdot x^i_k - p^i_k \geq v_i \cdot x^i_{k+1} - p^i_{k+1}, \qquad v_i \cdot x^i_k - p^i_k \geq v_i \cdot x^i_{k-1} - p^i_{k-1}.$$

Rearranging the inequalities, we get:

$$\frac{p^i_{k+1} - p^i_k}{x^i_{k+1} - x^i_k} \geq v_i \geq \frac{p^i_k - p^i_{k-1}}{x^i_k - x^i_{k-1}}.$$

The claim follows by inductively applying the claim for each two consecutive intervals. □

**Claim C.2** *The mechanism $M'$ the algorithm produces is universally truthful.*

**Proof:** Observe that after choosing $r$, $M'$ elicits bids, and then makes an allocation decision and a take-it-or-leave-it offer for each bidder, where the price for bidder $i$ does not depend on bidder $i$'s valuation but only on $v_{-i}$ and $r$ (by Claim C.1 if bidder $i$ is assigned the item then $v_i - p \geq 0$, otherwise $v_i - p \leq 0$). Notice that when a bidder has a zero profit from taking the item we can break ties arbitrarily without affecting truthfulness. Hence, for each $r$ the algorithm is truthful. Since we choose $r$ in a way that is independent of the bids, $M'$ is a universally truthful mechanism. □

**Claim C.3** *$M'$ implements $M$.*

**Proof:** Suppose that bidder $i$ with value $v_i$ is assigned $x$ and his payment is $p$ in $M$. By definition $(x,p)$ is equal to some $(x^i_k, p^i_k) \in \mathcal{R}^i_{v_{-i}}$. In the algorithm, it is easy to see that the probability that bidder $i$ is assigned the item is exactly $x$. The expected payment of bidder $i$ is:

$$\frac{p^i_1}{x^i_1} \cdot x^i_1 + \frac{p^i_2 - p^i_1}{x^i_2 - x^i_1} \cdot (x^i_2 - x^i_1) + \cdots + \frac{p^i_k - p^i_{k-1}}{x^i_k - x^i_{k-1}} \cdot (x^i_k - x^i_{k-1}) = p^k_i = p.$$

□

**Claim C.4** *$M'$ is feasible.*

**Proof:** Given valuations $(v_1, v_2)$, suppose that $M$ assigns fractions $x$ and $y$ to the two bidders, respectively, then by the feasibility of $M$, $x + y \leq 1$. Therefore, bidder 2 is assigned the item exactly when $1 - r \leq y$, i.e., $r \geq 1 - y \geq x$, which is the condition that bidder 1 is not assigned the item. This guarantees the feasibility of $M'$. □

We have proved correctness of the algorithm in Section 4.1. Now we extend it to a procedure for the case when $\mathcal{R}^i_{v_{-i}}$'s can be infinite.

Similarly to the algorithm presented in Section 4.1, the mechanism first chooses a number $r \in [0, 1]$ uniformly at random. For every value of $r$, we have a deterministic mechanism as follows:

1. Elicit a bid from each bidder. Let the bids be $(v_1, v_2)$.

2. If $r \leq M(v_1, v_2)_1$, allocate the item to bidder 1; if $1 - r < M(v_1, v_2)_2$, allocate the item to bidder 2.

3. By the characterization of individually rational truthful mechanisms in this setting, the payment of the bidder who gets the item is determined by the monotone allocation rule specified in the last step — if bidder 1 gets the item, he makes the payment $\inf\{v_1 \mid M(v_1, v_2)_1 > r\}$; if bidder 2 gets the item, he makes the payment $\inf\{v_2 \mid M(v_1, v_2)_2 > 1 - r\}$.

By similar arguments as in the previous claims, for every value of $r$ this deterministic mechanism is truthful and feasible. We therefore obtain a distribution over deterministic truthful mechanisms, i.e., a universally truthful mechanism. This mechanism implements the allocation rule of $M$, and by the uniqueness of payment scheme for individually rational truthful mechanisms, the payment of $M$ is implemented as well.

We note that the algorithm presented in Section 4.1 for the finite case is a special case of the procedure presented above. We presented it in an apparently different way simply to make the algorithm more explicit and to make it clear that its running time is polynomially bounded in that case.

This completes the proof of Theorem 4.3.

We further remark that for a truthful-in-expectation mechanism $M$ that is not IR and $p_0^i$ is not 0, we can still find a universally truthful mechanism that implements it. We simply change the labels to be $\frac{p_{k+1}^i + p_0^i - p_k^i}{x_{k+1}^i - x_k^i}$, and the rest of the argument (except Claim C.1) goes through.

## C.2 More on Implementability

**Theorem C.5** *In a single item auction for n bidders, if each bidder has at most two possible values then every truthful in expectation mechanism can be implemented by a universally truthful mechanism.*

**Proof:** Let bidder $i$'s two values be $\ell_i$ and $h_i$, with $\ell_i < h_i$. We construct a graph $G$ whose nodes are all possible bid vectors, i.e., each node $v$ is labeled $(\alpha_i)_i$, where $\alpha_i$ is from $\{\ell_i, h_i\}$. There is an edge between two nodes if their labels differ in exactly one coordinate. We then add some dummy nodes to $G$: for each node $v = (\alpha_i)_i$ where $|\{i \mid \alpha_i = h_i\}| = k$, add $k$ dummy nodes and identify each one with bidder $i$ for which $\alpha_i = h_i$ in $v$. Add an edge between $v$ and each of the dummy nodes. These dummy nodes have no other edges, and therefore have degree 1. Observe that $G$ is a bipartite graph.

We will show that every integral matchings in $G$ corresponds to a truthful deterministic mechanisms. Furthermore, we will show that a fractional matching corresponds to a truthful-in-expectation mechanisms. The theorem will then follow since it is known that a fractional matching can be "decomposed" to a probability distribution over integral matchings in a bipartite graph (see below).

Observe that every matching $M$ in $G$ corresponds to a deterministic truthful mechanism: for each bid vector, consider the corresponding node in $G$. If it is not matched to any other node, then allocate nothing; if it is matched to a dummy node, then allocate the item to bidder $i$ if the dummy node corresponds to $i$ (bidder $i$'s value must be $h_i$ in this case); otherwise, $v$ is matched to a node corresponding to $v'$, then allocate the item to bidder $i$ where $i$ is the coordinate where $v$ and $v'$ differs. The mechanism is feasible. To check truthfulness we show that the algorithm is monotone: if bidder $i$'s value is $\ell_i$ in $v$ and $i$ gets the item, then $v$ is matched to $v'$ and the two differ only in coordinate $i$; therefore, if, all other bids fixed, his value changes to $h_i$, he still gets the item.

The next observation we make is that any truthful-in-expectation mechanism $M$ corresponds to a fractional matching in $G$ in a similar way. Suppose at bid vector $v$, $M$ allocates a fraction $x$ of the item to bidder $i$. Let $v'$ be the bid vector that differs from $v$ only at coordinate $i$ and suppose $M$ allocates fraction $y$ of the item to $i$ at $v'$. If $v_i$ is $\ell_i$, then add $x$ fraction of the edge $(v, v')$ to the fractional matching. Otherwise add $x - y$ fraction of the edge from $v$ to the dummy node corresponding to the $h_i$ in $v$. We are guaranteed that $x - y$ is nonnegative by the monotonicity of $M$. We also know that the total weight of all edges incident to a node does not exceed 1 by the feasibility of $M$. Hence we get a fractional matching on $G$.

We note that the correspondences described in the preceding two paragraphs are inverse to each other. In other words, given any matching $M$ in $G$, if we first map it to the corresponding deterministic truthful mechanism, and then map the mechanism back as a matching in $G$, then we get $M$ back. Therefore, we have established a one-to-one correspondence between all deterministic truthful mechanisms and all matchings in $G$. Furthermore, this correspondence is "linear", in the sense that if a fractional matching in $G$ corresponds to a TIE mechanism $M$ and can be expressed as a convex combination of a set of matchings in $G$, then $M$ can be expressed as the convex combination of the deterministic truthful mechanisms corresponding to these matchings, with the same linear coefficients. Expressing a TIE mechanism as a convex combination of deterministic truthful mechanisms, however, is exactly what we mean by implementing a TIE mechanism using universally truthful mechanism. The problem then boils down to decomposing a fractional matching in a bipartite graph as a convex combination of matchings on the same graph — Theorem A.5 and Lemma A.6 say exactly that this can always be done, which concludes the proof. □

# D  A gap between truthful-in-expectation and deterministic truthful mechanisms

This section describes a 3-bidder distribution whose optimal truthful-in-expectation mechanism is not deterministic. We furthermore prove that the revenue of this mechanism exceeds the revenue of the optimal deterministic truthful mechanism by a factor of at least 1.001.

## D.1  Description of the example

The joint distribution of type profiles is supported on a 27-element subset of $\mathbb{R}^3$. Each column of the following matrix represents a type profile in the support of the distribution.

$$\begin{bmatrix} 9 & 9 & 9 & 9 & 10 & 11 & 9 & 10 & 11 \\ 9 & 9 & 10 & 11 & 9 & 9 & 11 & 9 & 10 \\ 10 & 11 & 9 & 9 & 9 & 9 & 10 & 11 & 9 \end{bmatrix}$$

Each of these nine type profiles has probability $\frac{4}{126}$. In addition, for each type profile obtained by taking a column of the above matrix and increasing exactly one of its entries to 20, the probability of that type profile is $\frac{5}{126}$. The number of such type profiles is 18, so their total probability is $18 \cdot (5/126) = 90/126$. Combined with the matrix columns themselves, whose probabilities sum up to $9 \cdot (4/126) = 36/126$, we see that the given probabilities sum up to 1, meaning that they constitute a valid joint distribution.

We assume that bidders are numbered $1, 2, 3$, and that bidder indices are interpreted modulo 3; for example, if $i = 3$ then bidder $i + 1$ refers to bidder 1. Let $(v_1, v_2, v_3)$ denote the profile of bids. An optimal truthful-in-expectation mechanism $\mathcal{M}_{TIE}$ can be described as follows.

1. If $\max\{v_1, v_2, v_3\} \geq 20$ then we give the item to the lowest-numbered bidder whose bid is at least 20.

2. Else, if $10 \leq \max\{v_1, v_2, v_3\} < 20$, then bidder $i$'s probability of winning is equal to $1/2$ if the other pair of bids $(v_{i-1}, v_{i+1})$ belongs to the following set of ordered pairs: $\{(9, 9), (9, 10), (11, 9)\}$. Otherwise, bidder $i$'s probability of winning is zero.

3. If $\max\{v_1, v_2, v_3\} < 10$ then no one wins.

A case analysis reveals that the probabilities never sum to more than 1, and that for every bidder $i$, for every profile of the other bids $v_{-i}$, the probability of winning is monotonically non-decreasing in $v_i$.

## D.2  A linear program that eliminates payment variables

We begin by modifying the linear program from Section 3 to eliminate the payment variables and obtain an equivalent linear program whose only variables represent the allocation function. The fact that this is possible is a manifestation of the Revenue Equivalence Theorem, but it can also be proven by a direct manipulation of the linear program.

Maximizing revenue in a truthful-in-expectation mechanism is equivalent to solving a linear program specified below, where subscript $i$ ranges over bidders (i.e. $i = 1, 2, 3$) and subscript $s$ ranges over "scenarios", i.e. bid profiles in the product of the supports of each bidder's distribution. Define a relation $RB(i, s, t)$ to mean that $s, t$ are two scenarios and that $t$ is obtained from $s$ by increasing (or preserving) player $i$'s bid while leaving the other players' bids fixed.

In the original LP, we have variables $x_{is}$ to denote the probability that bidder $i$ gets the item in scenario $s$. We replace these variables by $y_{is}$'s that represent increments in $x_{is}$ when bidder $i$'s bid increases (with the other bids fixed), so that for each $s$ and $i$, $x_{is} = \sum_{t:RB(i,s,t)} y_{it}$. Therefore, by requiring $y_{is}$ to be nonnegative for each $s$, we guarantee the monotonicity of the mechanism. The feasibility condition $\sum_i x_{is} \leq 1$ can be directly translated as $\sum_{i,t:RB(i,t,s)} y_{it} \leq 1$.

In order to have an LP for the maximum revenue, we still need to specify payments. In the following we show that, given $y_{is}$'s, we can express the maximum revenue achievable without introducing new variables. [10]

Suppose $v_{is}$ denotes the valuation of bidder $i$ in scenario $s$. When $v_{is}$ is 0, by individual rationality we know that the expected payment $p_{is}$ is 0. Let $s'$ be the scenario in which bidder $i$ has increased his bid to the next larger value in the support while all other bids remain unchanged. By truthfulness we have $x_{is'}v_{is'} - p_{is'} \geq x_{is}v_{is'} - p_{is}$, i.e., $p_{is'} \leq v_{is'}(x_{is'} - x_{is}) + p_{is} = v_{is'}y_{is'}$. Inductively, for any scenario $t$ we have that $p_{it} \leq \sum_{s:R(i,s,t)} v_{is}y_{is}$. Therefore, the revenue of the mechanism can be no larger than if we simply set the expected payments in scenario $t$ to be $\sum_{s:R(i,s,t)} v_{is}y_{is}$. It is easy to check that this payment scheme satisfies individual rationality conditions and the other truthful conditions. Therefore this payment scheme gives the maximum revenue under $y_{is}$'s, and we can express the revenue as

$$\sum_{s,i} \pi_s \sum_{t:RB(i,t,s)} v_{it} y_{it},$$

where $\pi_s$ denotes the probability of scenario $s$. [11]

To simplify notation, we define

$$\psi_{is} = v_{is} \cdot \left( \sum_{t:RB(i,s,t)} \pi_t \right),$$

and change the order of summations, the new objective function can be expressed as $\sum_{i,s} \psi_{is} y_{is}$.

Now we write out the full new LP, which we have shown to give the maximum revenue of truthful-in-expectation mechanisms:

$$\begin{aligned} \max \quad & \sum_{i,s} \psi_{is} y_{is} \\ \text{s.t.} \quad & \sum_{i,s:RB(i,s,t)} y_{is} \leq 1 \quad \forall t \\ & y_{is} \geq 0 \quad \forall i, s \end{aligned}$$

The dual is the following linear program.

$$\begin{aligned} \min \quad & \sum_t z_t \\ \text{s.t.} \quad & \sum_{t:RB(i,s,t)} z_t \geq \psi_{is} \quad \forall i, s \\ & z_t \geq 0 \quad \forall t \end{aligned}$$

---

[10]This is in essence a rederivation of a characterization result from Myerson in the discrete case.

[11]One way of making these payments ex post individually rational is to charge a random amount to player $i$. Conditional on $i$ winning in scenario $t$, the amount $v_{is}$ is charged with probability $y_{is}/x_{it}$ for every $s$ such that $RB(i, s, t)$.

## D.3 Analysis of the example

For the distribution specified in Section D.1, the optimal primal and dual solutions are specified as follows. For notational convenience, we use $a, b, c, d$ to denote $9, 10, 11, 20$, respectively. Thus, for example $y_{2,adc}$ would denote the variable $y_{is}$ for bidder $i = 2$ in scenario $s = (a, d, c) = (9, 20, 11)$. Also for convenience, we will only specify the nonzero $y_{is}$ and $z_t$ variables.

The primal solution has 27 nonzero values. Nine of them are:

$$y_{2,aab} = y_{3,aab} = y_{1,aba} = y_{2,aba} = y_{1,baa} = y_{3,baa} = y_{1,aac} = y_{2,caa} = y_{3,aca} = 1/2 \qquad (1)$$

In addition, for every scenario $s$ in which player $i$'s bid is 20, the primal variable $y_{is}$ is equal to $1 - (y_{it} + y_{iu} + y_{iv})$, where $t, u, v$ are obtained from $s$ by changing player $i$'s bid to $9, 10, 11$, respectively. Thus,

$$y_{1,daa} = y_{1,dac} = y_{1,dba} = y_{2,ada} = y_{2,cda} = y_{2,adb} = y_{3,aad} = y_{3,acd} = y_{3,bad} = 1/2$$
$$y_{1,dab} = y_{1,dca} = y_{1,dcb} = y_{2,bda} = y_{2,adc} = y_{2,bdc} = y_{3,abd} = y_{3,cad} = y_{3,cbd} = 1$$

The dual solution has 27 nonzero values, one for each scenario. Nine of them are:

$$z_{aab} = z_{aba} = z_{baa} = z_{aac} = z_{aca} = z_{caa} = \frac{15}{126}$$
$$z_{acb} = z_{bac} = z_{cba} = \frac{2}{126}$$

In addition, for every scenario $s$ in which one player's bid was raised to 20, the dual variable $z_s$ is equal to $\frac{100}{126}$.

It remains to check that these primal and dual solutions are feasible, that they satisfy both primal and dual complementary slackness, and that the primal solution $y$ is the unique solution satisfying complementary slackness with respect to $z$. (This last step is necessary in order to substantiate the claim that there is no deterministic truthful mechanism which gets as much revenue.)

Primal feasibility and primal complementary slackness are easy to check. To verify dual feasibility and dual complementary slackness, we need to check 192 constraints (3 players times 64 scenarios) and we do it using a five-case analysis.

**Case 1: Pairs $i, s$ such that at least one bid in $s$ is equal to 20.** The value of $\psi_{is}$ is equal to $k_{is} \cdot v_{is} \cdot \left(\frac{5}{126}\right)$, where $k_{is}$ denotes the number of scenarios $t$ such that $RB(i, s, t)$ holds. The value of $\sum_{t:RB(i,s,t)} z_t$ is $k_{is} \cdot \frac{100}{126}$. Since $v_{is} \leq 20$ for all $i, s$, we see that these constraints are always satisfied, and that the tight constraints are precisely those with $k_{is} = 0$ or $v_{is} = 20$.

**Case 2: All other pairs $i, s$ such that $y_{is} > 0$.** There are 9 remaining pairs $i, s$ such that $y_{is} > 0$, and they break up into 3 orbits under the action of the symmetry group that cyclically permutes the bidders' identities. It suffices to check one representative of each of these orbits, for example the constraints corresponding to primal variables $y_{2,aab}, y_{3,aab}, y_{1,aac}$. These constraints are

$$\psi_{2,aab} = z_{aab} + z_{acb} + z_{adb}$$
$$\psi_{3,aab} = z_{aab} + z_{aac} + z_{aad}$$
$$\psi_{1,aac} = z_{aac} + z_{bac} + z_{dac}$$

We now compute that

$$\psi_{2,aab} = 9 \cdot \left(\frac{4}{126} + \frac{4}{126} + \frac{5}{126}\right) = 9 \cdot \frac{4+4+5}{126} = \frac{117}{126}$$

$$z_{aab} + z_{acb} + z_{adb} = \frac{15}{126} + \frac{2}{126} + \frac{100}{126} = \frac{117}{126}$$

$$\psi_{3,aab} = 10 \cdot \left(\frac{4}{126} + \frac{4}{126} + \frac{5}{126}\right) = 10 \cdot \frac{4+4+5}{126} = \frac{130}{126}$$

$$z_{aab} + z_{aac} + z_{aad} = \frac{15}{126} + \frac{15}{126} + \frac{100}{126} = \frac{130}{126}$$

$$\psi_{1,aac} = 9 \cdot \left(\frac{4}{126} + \frac{4}{126} + \frac{5}{126}\right) = 9 \cdot \frac{4+4+5}{126} = \frac{117}{126}$$

$$z_{aac} + z_{bac} + z_{dac} = \frac{15}{126} + \frac{2}{126} + \frac{100}{126} = \frac{117}{126},$$

which confirms dual feasibility and complementary slackness in these cases.

**Case 3: All remaining pairs $(i,s)$ such that $\pi_s > 0$.** When all bids are less than 20, there are only nine scenarios $s$ such that $\pi_s > 0$, namely the nine columns of the matrix above. For each column of the matrix there are three choices of $i$, making for 27 pairs $(i,s)$ in total. However, 9 of these pairs were checked in Case 2 above. This leaves 18 pairs that form 6 orbits under the action of the cyclic symmetry group. The following list of 6 constraints contains one representative of each orbit.

$$\psi_{1,aab} \leq z_{aab} + z_{dab}$$
$$\psi_{2,aac} \leq z_{aac} + z_{adc}$$
$$\psi_{3,aac} \leq z_{aac} + z_{aad}$$
$$\psi_{1,acb} \leq z_{acb} + z_{dcb}$$
$$\psi_{2,acb} \leq z_{acb} + z_{adb}$$
$$\psi_{3,acb} \leq z_{acb} + z_{acd}$$

The left sides of these 6 constraints are computed as follows.

$$\psi_{1,aab} = 9 \cdot \left(\frac{4}{126} + \frac{5}{126}\right) = \frac{81}{126}$$

$$\psi_{2,aac} = 9 \cdot \left(\frac{4}{126} + \frac{5}{126}\right) = \frac{81}{126}$$

$$\psi_{3,aac} = 11 \cdot \left(\frac{4}{126} + \frac{5}{126}\right) = \frac{99}{126}$$

$$\psi_{1,acb} = 9 \cdot \left(\frac{4}{126} + \frac{5}{126}\right) = \frac{81}{126}$$

$$\psi_{2,acb} = 11 \cdot \left(\frac{4}{126} + \frac{5}{126}\right) = \frac{99}{126}$$

$$\psi_{3,acb} = 10 \cdot \left(\frac{4}{126} + \frac{5}{126}\right) = \frac{90}{126}$$

Note that the left side of each constraint is at most $\frac{99}{126}$. The right side of each constraint is at least $\frac{102}{126}$, because in each of the six constraints, the first term on the right side is at least $\frac{2}{126}$ and the second term is equal to $\frac{100}{126}$. Therefore, each of the six constraints holds, which completes Case 3.

**Case 4: Pairs $(i,s)$ such that $\pi_s = 0$ but $\psi_{is} > 0$.** If $\psi_{is} > 0$ then there is a scenario $t$ such that $\pi_t > 0$ and $RB(i,s,t)$ holds. Consider the least such $t$, i.e. the one obtained from $s$ by raising $i$'s bid by the minimum amount. Then $\psi_{it} = \psi_{is} \cdot (v_{it}/v_{is}) > \psi_{is}$. As $\pi_t > 0$, we have already verified the constraint $\sum_{u:RB(i,t,u)} z_u \geq \psi_{it}$ in one of the earlier cases. Now it follows that $\sum_{u:RB(i,s,u)} z_u > \psi_{is}$, i.e. the dual constraint indexed by $(i,s)$ holds and it is not tight.

**Case 5: Pairs $(i,s)$ such that $\psi_{is} = 0$.** In this case, it trivially holds that $\psi_{is} \leq \sum_{t:RB(i,s,t)} z_t$, with equality if and only if all of the dual variables in the sum on the right are zero.

We are now in a position to prove that every optimal primal solution satisfies (1). The dual variables $z_t$ are strictly positive for every $t \in \{aab, aba, baa, aac, aca, caa, acb, bac, cba\}$. Via complementary slackness, this gives nine equations that must be satisfied by every optimal primal solution.

$$\sum_{i,s:RB(i,s,t)} y_{is} = 1 \qquad \forall t \in \{aab, aba, baa, aac, aca, caa, acb, bac, cba\}. \tag{2}$$

Of the primal variables $y_{is}$ that participate in these nine linear equations, many of them are required, by complementary slackness, to take the value 0 at any optimal primal solution because the corresponding dual constraint is not tight. We can determine which of the relevant variables $y_{is}$ are allowed to take a nonzero value by reviewing the case analysis above. If $(i,s)$ satisfies $RB(i,s,t)$ for some $t \in \{aab, aba, baa, aac, aca, caa, acb, bac, cba\}$, then $\psi_{is} > 0$ and no bid in $s$ is equal to 20. This excludes Cases 5 and 1, respectively. None of the dual constraints in Cases 3 and 4 are tight, so none of the corresponding variables $y_{is}$ can take a nonzero value. This leaves Case 2, which corresponds to the nine variables occurring in (1). Rewriting the system of linear equations (2) in terms of these nine variables, we obtain the linear system

$$\begin{bmatrix} 1 & 1 & 0 & 0 & 0 & 0 & 0 & 0 & 0 \\ 0 & 1 & 1 & 0 & 0 & 0 & 0 & 0 & 0 \\ 0 & 0 & 1 & 1 & 0 & 0 & 0 & 0 & 0 \\ 0 & 0 & 0 & 1 & 1 & 0 & 0 & 0 & 0 \\ 0 & 0 & 0 & 0 & 1 & 1 & 0 & 0 & 0 \\ 0 & 0 & 0 & 0 & 0 & 1 & 1 & 0 & 0 \\ 0 & 0 & 0 & 0 & 0 & 0 & 1 & 1 & 0 \\ 0 & 0 & 0 & 0 & 0 & 0 & 0 & 1 & 1 \\ 1 & 0 & 0 & 0 & 0 & 0 & 0 & 0 & 1 \end{bmatrix} \begin{bmatrix} y_{2,aab} \\ y_{3,aab} \\ y_{1,aac} \\ y_{3,baa} \\ y_{1,baa} \\ y_{2,caa} \\ y_{1,aba} \\ y_{2,aba} \\ y_{3,aca} \end{bmatrix} = \begin{bmatrix} 1 \\ 1 \\ 1 \\ 1 \\ 1 \\ 1 \\ 1 \\ 1 \\ 1 \end{bmatrix} \tag{3}$$

The matrix on the left side is invertible, so the unique solution of the linear system is the one given in (1).

### D.4 Bounding the integrality gap

Going a step further, we can bound the integrality gap in this example by observing that integer solutions of the primal LP must satisfy an additional constraint. The key observation is that the constraints of the primal LP imply the relation

$$\begin{bmatrix} 1 & 1 & 0 & 0 & 0 & 0 & 0 & 0 & 0 \\ 0 & 1 & 1 & 0 & 0 & 0 & 0 & 0 & 0 \\ 0 & 0 & 1 & 1 & 0 & 0 & 0 & 0 & 0 \\ 0 & 0 & 0 & 1 & 1 & 0 & 0 & 0 & 0 \\ 0 & 0 & 0 & 0 & 1 & 1 & 0 & 0 & 0 \\ 0 & 0 & 0 & 0 & 0 & 1 & 1 & 0 & 0 \\ 0 & 0 & 0 & 0 & 0 & 0 & 1 & 1 & 0 \\ 0 & 0 & 0 & 0 & 0 & 0 & 0 & 1 & 1 \\ 1 & 0 & 0 & 0 & 0 & 0 & 0 & 0 & 1 \end{bmatrix} \begin{bmatrix} y_{2,aab} \\ y_{3,aab} \\ y_{1,aac} \\ y_{3,baa} \\ y_{1,baa} \\ y_{2,caa} \\ y_{1,aba} \\ y_{2,aba} \\ y_{3,aca} \end{bmatrix} \preceq \begin{bmatrix} 1 \\ 1 \\ 1 \\ 1 \\ 1 \\ 1 \\ 1 \\ 1 \\ 1 \end{bmatrix} \tag{4}$$

which is obtained by taking the $9 \times 9$ submatrix of the LP constraint matrix whose rows are indexed by scenarios $aab, aac, bac, baa, caa, cba, aba, aca, acb$, in that order, and whose columns are indexed by the nine variables that appear in the column vector on the left side of the inequality. Multiplying the left and right sides of (4) by the row vector $(\frac{1}{2}, \frac{1}{2}, \frac{1}{2}, \frac{1}{2}, \frac{1}{2}, \frac{1}{2}, \frac{1}{2}, \frac{1}{2}, \frac{1}{2})$, we obtain

$$y_{2,aab} + y_{3,aab} + y_{1,aac} + y_{3,baa} + y_{1,baa} + y_{2,caa} + y_{1,aba} + y_{2,aba} + y_{3,aca} \leq \frac{9}{2}.$$

Together with the constraints $y_{is} \in \mathbb{Z}$ for all $i, s$, this implies the stronger inequality

$$y_{2,aab} + y_{3,aab} + y_{1,aac} + y_{3,baa} + y_{1,baa} + y_{2,caa} + y_{1,aba} + y_{2,aba} + y_{3,aca} \leq 4. \tag{5}$$

If we add this constraint into the primal LP, it modifies the dual by adding an extra variable $\zeta$ that enters the dual objective function with coefficient 4, and that enters various dual constraints (those corresponding to primal variables occurring in (5)) with coefficient 1. If we now set

$$z_{aab} = z_{aba} = z_{baa} = z_{aac} = z_{aca} = z_{caa} = \frac{13}{126}$$

$$z_{acb} = z_{bac} = z_{cba} = 0$$

$$\zeta = \frac{4}{126}$$

and leave the remaining dual variables unchanged, it maintains dual feasibility. This can be verified by repeating the case analysis above. Cases 1,4,5 are unaffected by the change. In Case 2 the change in the $z$ variables reduces the right side of each constraint by $\frac{4}{126}$ but this is compensated by the introduction of $\zeta$ into the right side. In Case 3 the change in the $z$ variables reduces the right side of each constraint by $\frac{2}{126}$, but the analysis in Case 3 established that the right side of each constraint already exceeded the left side by at least $\frac{3}{126}$, so reducing the right side by $\frac{2}{126}$ maintains dual feasibility.

Consequently, the objective value of the new dual LP at this feasible solution constitutes an upper bound on the value of any integer primal solution. The new dual objective value is

$$4 \cdot \frac{4}{126} + 6 \cdot \frac{13}{126} + 18 \cdot \frac{100}{126} = \frac{1894}{126},$$

whereas the old dual objective value was

$$6 \cdot \frac{15}{126} + 3 \cdot \frac{2}{126} + 18 \cdot \frac{100}{126} = \frac{1896}{126}.$$

Thus, the revenue of the optimal truthful-in-expectation mechanism improves on the revenue of any deterministic truthful mechanism by at least a factor of $\frac{1896}{1894} > 1.001$.

**Remark:** In the above example, if we take out from the support all points that have a bid 20 in it, we get a smaller support of 9 points. From the randomized mechanism in the example we induce a mechanism on this smaller support. We claim that it is not implementable by any universally truthful mechanism on this support, i.e., it cannot be expressed a convex combination of deterministic truthful mechanisms.

To see this, consider any deterministic truthful mechanism and let $(y_{is})$ denote the corresponding LP solution. Since the values $y_{is}$ are all integers, we proved above that inequality (5) is satisfied. Our truthful-in-expectation mechanism does not satisfy this linear inequality (it has $y_{2,aab} + y_{3,aab} + y_{1,aac} + y_{3,baa} + y_{1,baa} + y_{2,caa} + y_{1,aba} + y_{2,aba} + y_{3,aca} = \frac{9}{2}$), and hence we have shown that even on this smaller support, this truthful-in-expectation mechanism cannot be written as a convex combination of deterministic truthful mechanisms.

Note that in this remark, we used only the fact $a < b < c$ and not the concrete values of $a, b$ and $c$. Therefore we have shown the following lemma, which is asserted in the second sentence of Theorem 4.7.

**Lemma D.1** *For three bidders, each bidder having only three possible values in the support, there is a truthful in expectation mechanism that cannot be implemented by any universally truthful mechanism.*